\documentclass[a4paper]{article}
\usepackage{amsmath}

\newcommand{\we}{\wedge}
\newcommand{\ot}{\otimes}
\newcommand{\ti}{\times}

\newcommand{\pa}{\partial}
\newcommand{\cD}{{\cal D}}

\newcommand{\cJ}{{\cal J}}
\newcommand{\cS}{{\cal S}}
\newcommand{\cN}{{\cal N}}

\newcommand{\cF}{{\cal F}}
\newcommand{\cO}{{\cal O}}
\newcommand{\cP}{{\cal P}}
\newcommand{\cR}{{\cal R}}

\newcommand{\cE}{{\cal E}}

\newcommand{\cH}{{\cal H}}
\newcommand{\cW}{{\cal W}}
\newcommand{\la}{\langle}
\newcommand{\ra}{\rangle}

\newcommand{\raa}{\rightarrow}

\newcommand{\R}{{\mathbf R}}
\newcommand{\C}{{\mathbf C}}
\newcommand{\D}{\text{d}}
\newcommand{\wh}{\widehat}
\newcommand{\wt}{\widetilde}

\newcommand{\ep}{\rule{1.5mm}{2.5mm}\\}
\mathchardef\za="710B  %\alpha
\mathchardef\zb="710C  %\beta
\mathchardef\zg="710D  %\gamma
\mathchardef\zd="710E  %\delta
\mathchardef\zve="710F %\epsilon
\mathchardef\zz="7110  %\zeta
\mathchardef\zh="7111  %\eta
\mathchardef\zvy="7112 %\theta
\mathchardef\zi="7113  %\iota
\mathchardef\zk="7114  %\kappa
\mathchardef\zl="7115  %\lambda
\mathchardef\zm="7116  %\mu
\mathchardef\zn="7117  %\nu
\mathchardef\zx="7118  %\xi
\mathchardef\zp="7119  %\pi
\mathchardef\zr="711A  %\rho
\mathchardef\zs="711B  %\sigma
\mathchardef\zt="711C  %\tau
\mathchardef\zu="711D  %\upsilon
\mathchardef\zvf="711E %\phi
\mathchardef\zq="711F  %\chi
\mathchardef\zc="7120  %\psi
\mathchardef\zw="7121  %\omega
\mathchardef\ze="7122  %\varepsilon
\mathchardef\zy="7123  %\vartheta
\mathchardef\zf="7124  %\varomega
\mathchardef\zvr="7125 %\varrho
\mathchardef\zvs="7126 %\varsigma
\mathchardef\zf="7127  %\varphi
\mathchardef\zG="7000  %\Gamma
\mathchardef\zD="7001  %\Delta
\mathchardef\zY="7002  %\Theta
\mathchardef\zL="7003  %\Lambda
\mathchardef\zX="7004  %\Xi
\mathchardef\zP="7005  %\Pi
\mathchardef\zS="7006  %\Sigma
\mathchardef\zU="7007  %\Upsilon
\mathchardef\zF="7008  %\Phi
\mathchardef\zW="700A  %\Omega

\newcommand{\be}{\begin{equation}}
\newcommand{\ee}{\end{equation}}

\newcommand{\bea}{\begin{eqnarray}}
\newcommand{\eea}{\end{eqnarray}}
\newcommand{\beas}{\begin{eqnarray*}}
\newcommand{\eeas}{\end{eqnarray*}}

\newtheorem{theo}{Theorem}
\newtheorem{prop}{Proposition}
\newtheorem{lem}{Lemma}
\newtheorem{cor}{Corollary}

\begin{document}

\title{Geometry of quantum systems: density states and entanglement}

\author{Janusz Grabowski\footnote{email: jagrab@impan.gov.pl} \\
\textit{Polish Academy of Sciences, Institute of Mathematics,} \\
\textit{\'Sniadeckich 8, P.O. Box 21, 00-956 Warsaw, Poland}          \\
\\
Marek Ku\'s\footnote{email: marek.kus@cft.edu.pl}\\
\textit{Center for Theoretical Physics, Polish Academy of Sciences,} \\
\textit{Aleja Lotnik{\'o}w 32/46, 02-668 Warszawa,
Poland} \\
\\
Giuseppe Marmo\footnote{email: marmo@na.infn.it}              \\
\textit{Dipartimento di Scienze Fisiche, Universit\`{a} ``Federico II'' di Napoli} \\
\textit{and Istituto Nazionale di Fisica Nucleare, Sezione di Napoli,} \\
\textit{Complesso Universitario di Monte Sant Angelo,} \\
\textit{Via Cintia, I-80126 Napoli, Italy} \\
}

\maketitle

\begin{abstract} Various problems concerning the geometry of the
space $u^*(\cH)$ of Hermitian operators on a Hilbert space $\cH$ are addressed.
In particular, we study the canonical Poisson and Riemann-Jordan tensors and
the corresponding foliations into K\"ahler submanifolds. It is also shown that
the space $\cD(\cH)$ of density states on an $n$-dimensional Hilbert space
$\cH$ is naturally a manifold stratified space with the stratification induced
by the the rank of the state. Thus the space $\cD^k(\cH)$ of rank-$k$ states,
$k=1,\dots,n$, is a smooth manifold of (real) dimension $2nk-k^2-1$ and this
stratification is maximal in the sense that every smooth curve in $\cD(\cH)$,
viewed as a subset of the dual $u^*(\cH)$ to the Lie algebra of the unitary
group $U(\cH)$, at every point must be tangent to the strata $\cD^k(\cH)$ it
crosses. For a quantum composite system, i.e.\ for a Hilbert space
decomposition $\cH=\cH^1\ot\cH^2$, an abstract criterion of entanglement is
proved.
\end{abstract}

\section{Introduction}
Dirac's approach to Quantum Mechanics uses a Hilbert space as a fundamental
object to start with, motivating the linear structure with the superposition
principle necessary to describe  phenomena like those of interference
\cite{Dirac}. Born's  probabilistic interpretation requires the use of a
Hermitian  inner product to deal with normalized states, therefore the physical
identification of states in the Hilbert space leads to the requirement that
(pure) states of a quantum mechanical system are described by elements of the
complex projective space (one-dimensional subspaces of a separable complex
Hilbert space $\cH$). By means of the Hermitian structure on $\cH$ it is
possible to define a binary product on the pure states $P\cH$
\cite{VonNeumann,Wigner,Weyl}. The physical interpretation of this binary
operation is given in terms of probability transition from one state to
another. On this space $P\cH$, bijective maps which preserve the transition
probability are necessarily projection of unitary or anti-unitary
transformations on the original Hilbert space, this statement is the main
content of Wigner's theorem \cite{Wigner1}.

More likely, due to this "equivalence" between the two descriptions (on $\cH$
and on $P\cH$), physicists have barely paid any attention to the geometrization
of Quantum Mechanics, i.e. to introduce a "tensorial description" in such a way
that non-linear coordinate transformations could be performed, notably
exception obviously do exist and we provide a partial list of references
\cite{all}. The recent great interest in the foundational aspects of Quantum
Mechanics motivated by the use of entanglement as a resource for quantum
information and quantum computing has boosted a more deep study of many
fundamental aspects, for instance the possibility to have a binary composition
of pure states without the use of the Hilbert space linear structure
\cite{Cirelli,MMSZ}, the possibility to have a non-linear Quantum Mechanics
\cite{Cantoni,Cirelli}, more generally the possibility to have non-linear
transformations among states.

The possibility of non-linear transformations may turn out to be quite useful
in the classification problem of separability and entanglement because these
properties are not preserved by taking linear combinations. Moreover, an
appropriate description of atomic phenomena involving polarization, spin
orientation and angular correlations, requires that we go beyond pure states in
the description of quantum systems. This larger family of states was introduced
by von Neumann as dual objects with respect to the quantum observables, they
constitute the set of density states and an early, physically motivated, review
was written by U.~Fano \cite{Fano}.

Again, for these states a proper mathematical setting is provided by the dual
space of the Lie algebra of the observables, with respect to the coadjoint
action of the unitary group. Density states emerge as elements of the coadjoint
orbits passing trough some special elements in the dual of the Cartan
subalgebra. The mathematical context of coadjoint orbits is quite well known to
those physicists involved with geometric quantization and the field was widely
studied in the seventies by Kostant, Kirillov, and Souriau \cite{KKS}.

Each coadjoint orbit bears a natural differential structure. Observe, however,
that the spectrum of the state does not change along the orbit of the unitary
action. From the point of view of quantum evolution it corresponds to the
situation of an isolated system, when all interactions with the environment are
negligible, so there is no dissipation and the evolution is unitary. In many
cases this is only a very exceptional situation, very rarely adequately
corresponding to the physical reality. On the other hand, it is \emph{a priori}
not clear that the whole set of density states, i.e.\ a union of coadjoint
orbits of the unitary action of different dimensionality, possesses a natural
differential structure. Exhibiting such a structure in terms of local
coordinates and/or \emph{via} a general geometric construction of a smooth
stratification of density states is thus of great interest when investigating
dissipative systems.

Density states form a convex subset of the set of Hermitian operators on
$\cal{H}$. Some properties of these convex body attracted recently an attention
\cite{Sommers}. It is thus legitimate to ask about "the shape" of the set of
density matrices, in particular about the smoothness properties of its
boundary. In the simplest case of the two-dimensional $\cal{H}$, the density
matrices form the three-dimensional unit ball with a smooth boundary - the
two-dimensional unit sphere comprising all pure states. But this situation is
exceptional - in higher dimensions the boundary does not consists exclusively
of pure states, it is in addition not smooth.

The space of  density states carries additional structures with respect to
those available on the space of coadjoint orbits of general Lie groups because
they are related to the unitary group and therefore  additional structures are
available. Moreover the need to consider composite quantum systems, tensor
products of the spaces associated with a choice of subsystems making up the
whole system, will bring up novel problems which will require further
investigations.

All these various considerations have convinced us that a review
of these mathematical aspects along with the identification of the
novel emerging problems may be useful to those people interested
in the application of quantum mechanics to quantum information and
are not at home with the geometrical background required. A recent
book by Chru\'sci\'nski and Jamio{\l}kowski \cite{Ch-Jam} deals
with geometrical aspects of quantum mechanics, these authors
however are primarily concerned with the application of these
methods to describe the geometric phase \cite{phase}. At this
point one should also point to the paper \cite{dittmann:92} in
which, in connection with geometric phase and parallel transport
along mixed states, the geometry of the manifold of density
matrices as a stratified space, was discussed along slightly
different lines than in the present paper (cf.\ Section 3 below).

In the present paper the Hilbert space $\cH$ will be assumed to be of finite
dimension $n$ in order to make the differential geometry expressible in local
coordinates classical. The reader will understand that passing to an
infinite-dimensional $\cH$ (i.e. a differential geometry of a Banach (or a
Hilbert) manifold) is straightforward, to this aim we will try to use
coordinate-free expressions, which serve in both cases, as much as possible.
The paper is organized as follows :

In section 2 we start with presenting the K\"ahler structure on the Hilbert
projective space $P\cH$ of pure states obtained from the standard Hermitian
product on $\cH$ via the momentum map associated with the Hamiltonian action of
the group $U(\cH)$ of unitary transformations of $\cH$. In this picture the
pure states form just an orbit in the dual space $u^*(\cH)$ of the unitary Lie
algebra $u(\cH)$ of the group $U(\cH)$. Because of the nondegeneracy of the
canonical invariant scalar product on $u(\cH)$ we have a canonical
identification of $u^*(\cH)$ with $u(\cH)$ which makes the geometry of
$u^*(\cH)$ very rich. We decided to interpret $u^*(\cH)$ as the space of
Hermitian operators on $\cH$ which makes possible to understand the density
states as a subset of $u^*(\cH)$.

Consequently, in sections 3 and 4 we present the density states as a convex
body $\cD(\cH)$ in $u^*(\cH)$ which is a family of some $U(\cH)$-orbits and, as
we will show later, also orbits of a particular action of the group ${GL}(\cH)$
of invertible complex linear operators on $\cH$. We show that $\cD(\cH)$ is
naturally a manifold stratified space with the stratification induced by the
the rank of the state. Thus the space $\cD^k(\cH)$ of rank-$k$ states,
$k=1,\dots,n$, is a smooth manifold of (real) dimension $2nk-k^2-1$ and this
stratification is maximal in the sense that every smooth curve in $\cD(\cH)$,
viewed as a subset of the dual $u^*(\cH)$ to the Lie algebra of the unitary
group $U(\cH)$, at every point must be tangent to the strata $\cD^k(\cH)$ it
crosses.

Section 5 is devoted to the geometry of $u^*(\cH)$, to a global description of
the K\"ahlerian structure of $U(\cH)$-orbits by means of the canonical Poisson
and Riemann-Jordan tensors. These K\"ahlerian structure are well-known in
algebraic geometry and can be easily generalized to analogous structures on
general flag manifolds. The point which should be stressed here is that the
geometry we develop is canonical, that it does not depend on the matrix form of
an operator and the $U(\cH)$-orbits are treated as a collection rather than
each orbit separately.

In the last section we investigate a Hilbert space decomposition
$\cH=\cH^1\ot\cH^2$ which is usually understood as corresponding to a quantum
composite system. We present an introduction to the problems of separability
and entanglement together with an abstract scheme for measurement of
entanglement.

Geometry of composite quantum systems was investigated in the literature from
several points of view. First, it is of importance to distinguish classes
of states which are equivalent under a restricted set of unitary
transformations (dubbed local transformations in the physical literature),
namely those which belong to the same orbit of $U(\cH^1)\ti U(\cH^2)$. From the
physical point of view all states on the same orbit contain an equal amount of
quantum correlations between the subsystems, i.e., these can not be influenced by
operations performed separately on each subsystem.

In order to characterize uniquely an orbit (i.e. a class of locally equivalent
states) one can try to find a complete set of $U(\cH^1)\ti U(\cH^2)$-invariant
functions on $\cD(\cH)$, such that the values of all functions at
$\rho\in\cD(\cH)$ characterize uniquely the orbit through $\rho$
\cite{schlienz:95,grassl:98}. The task can be effectively completed only for
low-dimensional systems - in fact, only in the the case dim$\cH^1$=dim$\cH^2=2$
the explicit results were found \cite{makhlin:02}. The same is true for
multipartite composite systems i.e. when
$\cH=\cH^1\otimes\cH^2\otimes\cdots\otimes\cH^K$. Here also the explicit
results are known for $K$ up to $4$ and dim$\cH^i=2$, $i=1,\ldots,4$
\cite{briand:03,luque:03}.

Other (partial) characterization of local orbits is provided by their
dimensions. These were investigated in \cite{linden:98,linden:99} and in
\cite{kus:01} all orbits of submaximal dimensionality in the case
dim$\cH^1$=dim$\cH^2=2$ were explicitly identified and enumerated. The similar
task of finding dimension of the local orbit through an arbitrary $\rho$  in
the case of higher dimensional systems was never achieved. A much modest goal
of determining dimensions and topology of local orbits stratifying the set of
rank one (pure) states $\cD^1(\cH^1\otimes\cH^2)$ was, however, completed for
arbitrary finite-dimensional $\cH^1$ and $\cH^2$ \cite{sinolecka:02}.

The sets of pure states in two- and three-partite systems with dim$\cH^i=2$ can
be identified with, respectively, unit seven- and fifteen- dimensional spheres
$\mathbf{S}^7$ and $\mathbf{S}^{15}$. In both cases there exist the Hopf
fibrations $\mathbf{S}^7\rightarrow\mathbf{S}^4$ and
$\mathbf{S}^{15}\rightarrow\mathbf{S}^8$ which were used to investigate the
geometry of pure states in \cite{moseri:01,moseri:03,levay:04,heydari:05}, whereas
multipartite pure states were treated in \cite{heydari:05a} using Segre variety.

Although in the present paper we limit ourselves to investigation
of two-partite composite system, we would like to point out recent
achievements in geometric characterization of entangled pure
states of multipartite systems. When investigating entanglement in
multicomponent system one aims at discriminating among different
classes of entanglement, defined as different equivalence classes
under appropriate group of transformations preserving entanglement
properties. The goal can be achieved by identification of the so
called entanglement monotones, i.e. measures of entanglement which
are invariant under considered transformations. Construction of
such invariants based on Pl\"ucker coordinates on Grassmannians,
naturally appearing when considering pure states of multicomponent
systems, were presented in \cite{levay:05} and \cite{heydari:05b}.
The geometry of three-qubit pure states entanglement was recently
investigated in \cite{levay:05a}, where geometric description of
different classes of entanglement was given in terms of
submanifolds of the so-called Klein quadric - a special quadric
embedded in the five-dimensional complex projective space.

Of special interest is the set of separable states (defined in Sec.\ 6), as
those which, from the physical point of view, do not carry any quantum
correlations. From the construction they form a convex subset in
$\cD(\cH^1\otimes\cH^2)$. Only in the case of dim$\cH^1$=dim$\cH^2=2$
and dim$\cH^1=2$ and dim$\cH^2=3$ (or {\it vice versa}) there exist effective
criteria which allow to discriminate separable and nonseparable (entangled)
states. As a consequence only in these low-dimensional case one can relative
easily investigate the geometry of the boundary of the set of separable states
\cite{shi:01}.

\section{K\"ahler structure on the Hilbert projective space}
Let $\cH$ be an $n$-dimensional Hilbert space with the Hermitian product $\la
x,y\ra_\cH$ being, by convention, $\C$-linear with respect to $y$ and
anti-linear with respect to $x$. The unitary group $U(\cH)$ acts on $\cH$
preserving the Hermitian product and it consists of those complex linear
operators $A\in gl(\cH)$ on $\cH$ which satisfy $AA^\dag=I$, where $A^\dag$ is
the Hermitian conjugate of $A$, i.e.
$$
\la Ax,y\ra_\cH=\la x,A^\dag y\ra_\cH.
$$
The geometric approach to Quantum Mechanics is based on considering the
realification $\cH_\R$ of $\cH$ as a K\"ahler manifold $(\cH_\R,J,g,\zw)$ with
canonical structures: a complex structure $J:\text{T}\cH_\R\raa \text{T}\cH_\R$,
a Riemannian metric $g$, and a symplectic form $\zw$. The latter come from the
real and the imaginary parts of the Hermitian product, respectively,
$g=\Re(\la\cdot,\cdot\ra_\cH)$, $\zw=\Im(\la\cdot,\cdot\ra_\cH)$. After the
obvious identification of the tangent bundle $\text{T}\cH_\R$ with
$\cH_\R\ti\cH_\R$, all these structures are constant structures induced from
$\cH$:
$$J(x)=i\cdot x,\qquad g(x,y)+i\cdot\zw(x,y)=\la x,y\ra_\cH.$$
We have obvious identities
$$J^2=-I,\quad \zw(x,Jy)=g(x,y),\quad g(Jx,Jy)=g(x,y),\quad
\zw(Jx,Jy)=\zw(x,y).$$
The tensors $g$ and $\zw$ being non-degenerate have their inverses: the
contravariant metric tensor $G=g^{-1}$ and the Poisson tensor $\zW=\zw^{-1}$.
They form together a Hermitian product
$$\la\za,\zb\ra_{\cH^*}=G(\za,\zb)+i\cdot\zW(\za,\zb)$$
on the dual real Hilbert space $\cH^*_\R$ equipped with the dual complex
structure $J^*$. Using the identification of $\cH^*_\R$ with $\cH_\R$ via the
metric tensor $g$, the latter can be interpreted as a contravariant complex
tensor on $\cH_\R$. This tensor induces two real brackets of smooth functions
on $\cH_\R$: $\{ f,h\}_g=G(\D f,\D h)$ and $\{ f,h\}_\zw=\zW(\D f,\D h)$. The
first one is the `Riemann-Jordan' bracket associated with the contravariant
version of the metric tensor $g$ and the second is just the symplectic Poisson
bracket associated with $\zw$. Of course both brackets can be extended to
complex functions by complex linearity and give rise to the `total' bracket
\be\{ f,h\}_\cH=\la \D f,\D h\ra_{\cH^*}=\{ f,h\}_g+ i\cdot\{ f,h\}_\zw. \ee

Fixing an orthonormal basis $(e_k)$ of $\cH$ allows us to identify the
Hermitian product $\la x,y\ra_\cH$ on $\cH$ with the canonical Hermitian
product on $\C^n$
\be\label{her}\la a,b\ra_{\C^n}=\overline{a_k}{b_k}\ee
(we use the convention of summation on repeated indices), the group $U(\cH)$ of
unitary transformations of $\cH$ with $U(n)$, its Lie algebra $u(\cH)$ with
$u(n)$, etc. In this picture $(a_{jk})^\dag=(\overline{a_{kj}})$ and $(T^\dag
T)_{jk}=\la\za_j,\za_k\ra$, where $\za_k=(t_{jk})\in\C^n$ are columns of the
matrix $T=(t_{jk})$. The choice of the basis induces (global) coordinates
$(q_k,p_k)$, $k=1,\dots,n$, on $\cH_\R$ by
$$\la e_k,x\ra_\cH=(q_k+i\cdot p_k)(x)$$
in which $\pa_{q_k}$ is represented by $e_k$ and $\pa_{p_k}$ by
$i\cdot e_k$. Hence the complex structure reads
$$J=\pa_{p_k}\ot\D q_k-\pa_{q_k}\ot\D p_k,$$
the Riemannian tensor
$$g=(\D q_k\ot\D q_k+\D p_k\ot\D p_k)=\frac{1}{2}
(\D q_k\vee\D g_k+\D p_k\vee\D p_k)$$ and the symplectic form
$$\zw=\D q_k\we \D p_k,$$
where $x\vee y=x\ot y+y\ot x$ is the symmetric, and $x\we y=x\ot
y-y\ot x$ is the wedge product. In complex coordinates
$z_k=q_k+i\cdot p_k$ one can write the Hermitian product as the
complex tensor
$$\la\cdot,\cdot\ra_\cH=\D\overline{z}_k\ot \D z_k.$$
The contravariant tensor $G+i\cdot\zW$ has the form
$$G+i\cdot\zW=(\pa_{q_k}\ot\pa_{q_k}+\pa_{p_k}\ot\pa_{p_k})+
i\cdot(\pa_{q_k}\ot\pa_{p_k}-\pa_{p_k}\ot\pa_{q_k})$$ or, in
complex coordinates,
$$G+i\cdot\zW=(\pa_{q_k}-i\cdot\pa_{p_k})\ot(\pa_{q_k}+i\cdot\pa_{p_k})
=4\pa_{z_k}\ot\pa_{\overline{z}_k}.$$ In other words, $$\{
f,h\}_g=\frac{\pa f}{\pa q_k}\frac{\pa h}{\pa q_k}+ \frac{\pa
f}{\pa p_k}\frac{\pa h}{\pa p_k},$$
$$\{ f,h\}_\zw=\frac{\pa f}{\pa q_k}\frac{\pa h}{\pa p_k}-
\frac{\pa f}{\pa p_k}\frac{\pa h}{\pa q_k},$$ and
$$\{ f,h\}_\cH=4\frac{\pa f}{\pa {z}_k}\frac{\pa h}{\pa
\overline{z}_k}.$$

\medskip Every complex linear operator $A\in gl(\cH)$ on $\cH$
induces the quadratic function
$$f_A(x)=\frac{1}{2}\la x,Ax\ra_\cH.$$
The function $f_A$ is real if and only if $A$ is Hermitian,
$A=A^\dag$.

\smallskip\noindent
One important convention we want to introduce is that we will
identify the space of Hermitian operators $A=A^\dag$ with the dual
$u^*(\cH)$ of the (real) Lie algebra $u(\cH)$, according to the
pairing between Hermitian $A\in u^*(\cH)$ and anti-Hermitian $T\in
u(\cH)$ operators
$$\la A,T\ra=\frac{i}{2}\cdot\text{Tr}(A T).$$
The multiplication by $i$ establishes further a vector space
isomorphism $u(\cH)\ni T\mapsto iT\in u^*(\cH)$ which identifies
the adjoint and the coadjoint action of the group $U(\cH)$,
$\text{Ad}_U(T)=U T U^\dag$. Under this isomorphism $u^*(\cH)$
becomes a Lie algebra with the Lie bracket
$[A,B]=\frac{1}{i}[A,B]_-$, where $[A,B]_-=AB-BA$ is the
commutator bracket, equipped additionally with the scalar product
\be\label{metric1}\la A,B\ra_{u^*}=\frac{1}{2}\text{Tr}(AB)
\ee
and an additional algebraic operation, the Jordan product
$[A,B]_+=AB+BA$. The scalar product is invariant with respect to
both: the Lie bracket and the Jordan product (or bracket)
\bea\label{invariant1}
\la[A,\zx],B\ra_{u^*(\cH)}&=&\la A,[\zx,B]\ra_{u^*(\cH)},\\
\label{invariant2} \la[A,\zx]_+,B\ra_{u^*(\cH)}&=&\la
A,[\zx,B]_+\ra_{u^*(\cH)}.
\eea
and it identifies once more $u^*(\cH)$ with its dual,
$$u^*(\cH)\ni A\mapsto\widehat{A}=\frac{1}{i}A\in u(\cH),$$
so vectors with covectors. Under this identification the metric
(\ref{metric1}) correspond to the invariant metric
\be\label{metric2}\la
\wh{A},\wh{B}\ra_{u}=\frac{1}{2}\text{Tr}(AB)
\ee
on $u(\cH)$ which can be viewed also as a contravariant metric on
$u^*(\cH)$.

\smallskip \noindent
For a (real) smooth function $f$ on $\cH_\R$ let us denote by
$grad_f$ and $Ham_f$ the gradient and the Hamiltonian vector field
associated with $f$ and the Riemannian and the symplectic tensor,
respectively. In other words, $g(\cdot,grad_f)=\D f$ and
$\zw(\cdot,{Ham}_f)=\D f$ or $grad_f=G(\D f,\cdot)$ and
$Ham_f=\zW(\D f,\cdot)$. Note that any $A\in gl(\cH)$ induces a
linear vector field $\widetilde{A}$ on $\cH$ by
$\widetilde{A}(x)=Ax$.
\begin{lem}For Hermitian $A$ we have
$$grad_{f_A}=\widetilde{A}\quad \text{and}\quad
Ham_{f_A}=\widetilde{i\,A}.$$
\end{lem}
{\it Proof.} If $\la\cdot,\cdot\ra$ denotes the pairing between
vectors and covectors then
\beas \la\D f_A(x),y\ra &=&\frac{1}{2}(\la y,Ax\ra_\cH+\la
x,Ay\ra_\cH)=\Re(\la y,Ax\ra_\cH)\\
&=& g(a,Ax)=\zw(y,iAx). \eeas \ep

\begin{cor} For all $A,B\in gl(\cH)$ we have
\be\{ f_A,f_B\}_\cH=f_{2AB}.
\ee In particular,
\bea\label{br1}
\{ f_A,f_B\}_g&=&f_{AB+BA},\\
\label{br2}\{ f_A,f_B\}_\zw&=&f_{-i(AB-BA)}. \eea
\end{cor}
{\it Proof.} For Hermitian $A,B$ we have
\beas\{ f_A,f_B\}_\cH(x)&=&g(grad_A(x),grad_B(x))+
i\cdot\zw(Ham_f(x),Ham_g(x))\\
&=&g(Ax,Bx)+i\cdot\zw(iAx,iBx)=\la Ax,Bx\ra_\cH=\la
x,ABx\ra_\cH=2f_{AB}(x).
\eeas
But $2AB=(AB+BA)+i(-i(AB-BA))$, where $AB+BA=[A,B]_+$ and
$-i(AB-BA)=-i[A,B]_-$ are Hermitian, thus $f_{[A,B]_+}$ and
$f_{-i[A,B]_-}$ are real, so the thesis holds for Hermitian $A,B$.
For general $A,B$ it follows by complex linearity. \ep

The unitary action of $U(\cH)$ on $\cH$ is in particular
Hamiltonian and induces a momentum map $\zm:\cH_\R\raa u^*(\cH)$.
The fundamental vector field associated with $\frac{1}{i}A\in
u(\cH)$, where $A\in u^*(\cH)$ is Hermitian, reads
$\widetilde{iA}$, since
$$\frac{\D}{\D t}\mid_{t=0}\exp{(-\frac{t}{i}A)}(x)=iA(x).$$ The
Hamiltonian of $\widetilde {iA}$ is $f_A$, so the momentum map is
defined by $$\la\zm(x),\frac{1}{i}A\ra=f_A(x)=\frac{1}{2}\la
x,Ax\ra_\cH.$$ But by our convention
$$\la\zm(x),\frac{1}{i}A\ra=\frac{i}{2}\text{Tr}(\zm(x)\frac{1}{i}A)=
\frac{1}{2}\text{Tr}(\zm(x)A),$$ so that $\text{Tr}(\zm(x)A)=\la
x,Ax\ra_\cH$ and finally, in the Dirac notation,
\be\zm(x)=\mid x\ra\la x\!\mid.
\ee

Note that for $A$ being Hermitian $f_A$ is the pullback
$f_A=\zm^*(\widehat{A})=\widehat{A}\circ\zm$, where
$\widehat{A}=\la A,\cdot\ra_{u^*}=\frac{1}{i}A\in u(\cH)$. The
linear functions $\widehat{A}$ generate $\mathrm{T}^*u^*(\cH)$, so that
(\ref{br1}) and (\ref{br2}) mean that the momentum map $\zm$
relates contravariant tensors $G$ and $\zW$ on $\cH$,
respectively, with the linear contravariant tensors $R$ and $\zL$
on $u^*(\cH)$ corresponding to the Jordan and Lie bracket,
respectively. The Riemann-Jordan tensor $R$, defined in the
obvious way,
\be
R(\zx)(\wh{A},\wh{B})=\la\zx,[A,B]_+\ra_{u^*}=\frac{1}{2}\text{Tr}
(\zx(AB+BA)),
\ee
is symmetric and the tensor
\be
\zL(\zx)(\wh{A},\wh{B})=\la\zx,[A,B]\ra_{u^*}=\frac{1}{2i}\text{Tr}
(\zx(AB-BA)),
\ee
is the canonical Kostant-Kirillov-Souriau Poisson tensor on
$u^*(\cH)$. They form together the complex tensor
\be(R+i\cdot\zL)(\zx)(\wh{A},\wh{B})=2\la\zx,AB\ra_{u^*}=
\text{Tr}(\zx AB)
\ee
and the momentum map relates this tensor with the dual Hermitian
product:
\be \zm_*(G+i\cdot\zW)=R+i\cdot\zL.
\ee
%tu
{\bf Example.} For $\cH=\C^2$ consider an orthonormal basis in
$u^*(2)$ consisting of
$$U=\left(\begin{array}{cc}1&0\\ 0&1\end{array}\right),\quad
X=\left(\begin{array}{cc}1&0\\ 0&-1\end{array}\right),\quad
Y=\left(\begin{array}{cc}0&1\\1&0\end{array}\right),\quad
Z=\left(\begin{array}{cc}0&i\\-i&0\end{array}\right)$$ and the
associated coordinates $u,x,y,z$, where
$u(A)=\frac{1}{2}\text{Tr(UA)}$, etc. In these coordinates the
Poisson tensor reads
$$\zL=2(z\pa_x\we\pa_y+x\pa_y\we\pa_z+y\pa_z\we\pa_x)$$
and the Riemann-Jordan tensor reads
$$R=\pa_u\vee(2x\pa_x+2y\pa_y+2z\pa_z)+u(\pa_u\vee\pa_u+\pa_x\vee\pa_x
+\pa_y\vee\pa_y+\pa_z\vee\pa_z).$$ The rank of $\zL(u,x,y,z)$ is 0
if $x^2+y^2+z^2=0$ and 2 if $x^2+y^2+z^2>0$. The rank of
$R(u,x,y,z)$ is 0 at $(u,x,y,z)=0$, it is 2 for $u=0$ and
$x^2+y^2+z^2>0$, it is 3 for $x^2+y^2+z^2=u^2>0$, and it is 4 for
$x^2+y^2+z^2\ne u^2>0$.

\bigskip%\noindent
The image $\zm(\cH\setminus\{ 0\})$ is the cone
$$\cP^1(\cH)=\{\mid\!x\ra\la
x\!\!\mid: x\ne 0\}$$ of non-negatively defined Hermitian
operators $\zx=\mid\!x\ra\la x\!\!\mid$ of rank 1. The operator
$\zx$ is proportional to the 1-dimensional projection
$\zx/\Vert\zx\Vert$, so $\zx^2=\Vert\zx\Vert\zx$, where
$\Vert\zx\Vert=\Vert x\Vert^2$ is the operator norm of $\zx$. The
manifold $\cP^1(\cH)$ is foliated by $U(\cH)$-coadjoint orbits
being complex projective spaces $\cD^1_r(\cH)=\{ \mid\!x\ra\la
x\!\!\mid:\Vert x\Vert=r\}$, $r>0$. In particular, the momentum
map image of the (2n-1)-dimensional sphere $S_\cH=\{ x\in\cH:\Vert
x\Vert^2=\la x,x\ra_\cH=1\}$ is the complex projective space
$\cD^1(\cH)=\{ \mid\!x\ra\la x\!\!\mid :\Vert x\Vert=1\}$ of
1-dimensional projectors.

The coadjoint orbits ${\cO}$ in $u^*(\cH)$ possess
canonical symplectic forms $\zh^{{\cO}}$ which build together the
Poisson structure $\zL$. These forms, as the inverses of
$\zL_{\mid{\cO}}$, are characterized by
\be \zh^{{\cO}}_\zx([A,\zx],[B,\zx])=\la[A,\zx],B\ra_{u^*(\cH)}
=-\la\zx,[A,B]\ra_{u^*}.
\ee
Indeed, the vectors $[A,\zx]=\frac{1}{i}[A,\zx]_-\ $ form the
tangent space of the $U(\cH)$-orbit through ${{\cO}}$ and
$\zh^{{\cO}}$ is the inverse of $\zL_{\mid{{\cO}}}$. Due to
invariance of the scalar product on $u^*(\cH)$:
\be\zL_\zx(\wh{A},\wh{B})=\la\zx,[A,B]\ra_{u^*(\cH)}=
\la[\zx,A],B\ra_{u^*(\cH)}=\la[\zx,A],\wh{B}\ra.
\ee
Hence $\#\zL_\zx(\wh{A})=[\zx,A]$ and
$$\#\zh^r_\zx([\zx,A])=(\#\zL_\zx)^{-1}([\zx,A])=\wh{A},$$
so
\beas\zh^r_\zx([A,\zx],[B,\zx])&=&\la\#\zh_\zx^r([A,\zx]),[B,\zx]\ra
=-\la\wh{A},[B,\zx]\ra\\
&=&-\la A,[B,\zx]\ra_{u^*(\cH)}=-\la\zx,[A,B]\ra_{u^*(\cH)}.
\eeas
The image $\# R(\text{T}^*u^*(\cH))$ of the tensor $R$ is not an
involutive (generalized)distribution, so its inverse $\zs=R^{-1}$
can be understood only as a `partial' covariant tensor on
$u^*(\cH)$, i.e. as a `partial symmetric 2-form' which at $\zx\in
u^*(\cH)$ is defined only on vectors from $\#
R_\zx(\text{T}_\zx^*u^*(\cH))$. There is a completely analogous
characterization of the tensor $\zs$ to that of $\zh$. Both
characterizations we can summarize as follows.
\begin{prop}\label{p2}
\begin{description}
\item{(a)} The symplectic form $\zh^{{\cO}}$ on the $U(\cH)$-orbit
${{\cO}}$ is characterized by
\be\label{sym} \zh^{{\cO}}_\zx([A,\zx],[B,\zx])=\la[A,\zx],B\ra_{u^*(\cH)}
=-\la\zx,[A,B]\ra_{u^*},
\ee
where $A,B\in u^*(\cH)$ are arbitrary Hermitian operators.

\item{(b)} The `partial tensor' $\zs$ on $u^*(\cH)$ is
characterized by
\be\label{1a}
\zs_\zx([A,\zx]_+,[B,\zx]_+)=\la[A,\zx]_+,B\ra_{u^*(\cH)}
=\la\zx,[A,B]_+\ra_{u^*}.
\ee
where $A,B\in u^*(\cH)$ are arbitrary Hermitian operators.
\end{description}
\end{prop}
Let us observe that the `partial tensor' $\zs$, when restricted to
any $\cD^1_r(\cH)$, induces a Riemannian structure $\zs^r$ which,
together with the symplectic structure $\zh^r=\zh^{\cD^1_r(\cH)}$,
induces a K\"ahler structure.
\begin{prop}\begin{description}
\item{(a)} The tensor $\zs^r$ being the restriction of the
partial tensor $\zs$ to the $U(\cH)$-orbit $\cD^1_r(\cH)$
through $\zx=\zm(x)$, $r^2=\Vert\zx\Vert$, is proportional to the
original scalar product on $u^*(\cH)$:
\be\label{22}
\zs^r_\zx([A,\zx],[B,\zx])=\frac{1}{\Vert\zx\Vert}
\la[A,\zx],[B,\zx]\ra_{u^*(\cH)}.
\ee
\item{(b)} The $(1,1)$-tensor $\cJ$ on $\cP^1(\cH)$,
$\cJ_\zx(A)=\frac{1}{\Vert\zx\Vert}[A,\zx]$, satisfies $\cJ^3=-\cJ$
and induces a complex structure $^r\!\cJ$ on every $\cD^1_r(\cH)$.
Moreover,
\be\label{x}\zh^r_\zx([A,\zx],^r\!\cJ_\zx([B,\zx]))=\zs^r_\zx([A,\zx],[B,\zx]),
\ee
and
\be\label{y}\zh^r_\zx(^r\!\cJ_\zx([A,\zx]),^r\!\cJ_\zx([B,\zx]))
=\zh^r_\zx([A,\zx],[B,\zx]),
\ee
i.e. $(\cD^1_r(\cH),^r\!\cJ,\zs^r,\zh^r)$ is a K\"ahler manifold
for each $r>0$.
\end{description}
\end{prop}
{\it Proof.} Observe first that, due to the Leibniz rule,
$$[A,\zx]=\frac{1}{\Vert\zx\Vert}[A,\zx^2]= \frac{1}{\Vert\zx\Vert}
[[A,\zx],\zx]_+.$$
Then, in view of (\ref{1a}),
$$\zs^r_\zx([A,\zx],[B,\zx])=\frac{1}{\Vert\zx\Vert^2}\la\zx,
[[A,\zx],[B,\zx]]_+\ra_{u^*(\cH)}= \frac{1}{2\Vert\zx\Vert^2}
\text{Tr}(\zx\circ[[A,\zx],[B,\zx]]_+).$$
But
\beas\text{Tr}(\zx\circ[[A,\zx],[B,\zx]]_+)&=&
\text{Tr}(\zx\circ[A,\zx]\circ [B,\zx]+\zx\circ[B,\zx]\circ
[A,\zx])\\
&=&\text{Tr}([A,\zx^2]\circ
[B,\zx]-[A,\zx]\circ\zx\circ[B,\zx]+\zx\circ[B,\zx]\circ
[A,\zx])\\
&=& \text{Tr}([A,\zx^2]\circ [B,\zx])
=\Vert\zx\Vert\text{Tr}([A,\zx]\circ [B,\zx])\\
&=& 2\Vert\zx\Vert\la[A,\zx],[B,\zx]\ra_{u^*(\cH)},
\eeas
that proves (\ref{22}).

To prove that $\cJ$ is a complex structure on every orbit, let us
recall that $\zx^2=\Vert\zx\Vert\zx$. Passing to
$\zx'=\zx/\Vert\zx\Vert$ if necessary, we can assume for all the
further calculations that $\Vert\zx\Vert=1$ so that
$\cJ_\zx(A)=[A,\zx]$. Hence,
\be\label{u}[[[A,\zx],\zx],\zx]=-\frac{1}{i}[(A\zx^2-2\zx A\zx+\zx^2
A),\zx]_-=-\frac{1}{i}[(A\zx^3-\zx^3A]=-[A,\zx]
\ee
and (\ref{22}) follows. Moreover, since vectors $[A,\zx]$ form the
tangent space $\text{T}_\zx\cD^1_r(\cH)$, (\ref{u}) shows that
$\cJ$ reduced to $\cD^1_r(\cH)$ is an almost-complex structure
$^r\!\cJ$. We shall show that the Nijenhuis torsion of $^r\!\cJ$
vanishes, so the structure is integrable. To do this, we must show
that the distribution in the complexified tangent bundle
$\text{T}\cD^1_r(\cH)\ot\C$ which corresponds to eigenvectors of
complexified $^r\!\cJ$ with the eigenvalue $i$ is involutive. But
this distribution is generated by complex vector fields
$\overline{T}$ for $T\in gl(\cH)$, where $\overline{T}(\zx)=\zx
T(1-\zx)$. Indeed,
$$\cJ_\zx(\zx T(1-\zx))=[\zx T(1-\zx),\zx]=\frac{1}{i}
(\zx T(1-\zx)\zx-\zx^2 T(1-\zx))= i\cdot\zx T(1-\zx)$$ and this is
a generating set due to the decomposition
$$T=(\zx T\zx+(1-\zx)T(1-\zx))+(1-\zx)T\zx+\zx T(1-\zx)$$ into
eigenvectors of $\cJ$ with eigenvalues 0, $-i$, and $i$,
respectively. The bracket of vector fields
$[\overline{T}_1,\overline{T}_2]_{vf}$ reads
\beas[\overline{T}_1,\overline{T}_2]_{vf}(\zx)&=&
\zx T_1(1-\zx)T_2(1-\zx)-\zx T_2\zx T_1(1-\zx) -\zx
T_2(1-\zx)T_1(1-\zx)\\&+&\zx T_1\zx T_1(1-\zx)
=\zx(T_1T_2-T_2T_1)(1-\zx)=\overline{([T_1,T_2]_-)}\eeas that
proves involutivity.

Finally, it is sufficient to combine (\ref{sym}) and (\ref{22}) to
get (\ref{x}). Then
\beas\zh^r_\zx(^r\!\cJ_\zx([A,\zx]),^r\!\cJ_\zx([B,\zx]))&=&
\zs^r_\zx(^r\!\cJ_\zx([A,\zx]),[B,\zx])=
\la^r\!\cJ_\zx([A,\zx]),[B,\zx]\ra_{u^*(\cH)}\\
&=&\la[[A,\zx],\zx],[B,\zx]\ra_{u^*(\cH)}=
-\la[[[A,\zx],\zx],\zx],B\ra_{u^*(\cH)}\\
&=&\la[[A,\zx],B\ra_{u^*(\cH)}= \zh^r_\zx([A,\zx],[B,\zx]),
\eeas
that proves (\ref{y}). \ep

\begin{prop} There is an identification of the orthogonal
complement of the vector $x\in\cH$ with the tangent space to the
$U(\cH)$-orbit through $\zx=\zm(x)$ in $u^*(\cH)$. For
$y,y'\in\cH$ orthogonal to $x$ with respect to the Hermitian
product, the vectors $(\zm_*)_x(y),(\zm_*)_x(y')$ are tangent to
the orbit through $\zx$ and
\bea\label{a}
\zs^r_\zx((\zm_*)_x(y),(\zm_*)_x(y'))&=&g(y,y'),\\
\label{b}\zh^r_\zx(,(\zm_*)_x(y'))&=&\zw(y,y'),\\
\label{c}^r\!\cJ_\zm(x)((\zm_*)_x(y))&=&(\zm_*)_x(Jy).
\eea
\end{prop}
{\it Proof.} Since
$$(\zm_*)_x(y)=P^x_y=\mid\!y\ra\la x\!\!\mid+\mid\!x\ra\la y\!\!\mid,$$
can be written as $P^x_y=[A_y,\zx]$, where $A_y$ is a Hermitian
operator such that $Ax=iy$ and $Ay=-i\frac{\Vert y\Vert^2}{\Vert
x\Vert^2}x$, the operators $P^x_y,P^x_{y'}$, viewed as vectors in
$u^*(\cH)$, are tangent to the orbit through $\zx$. Then, due to
(\ref{2}),
\beas\zs^r_\zx(P^x_y,P^x_{y'})&=&\frac{1}{2\Vert
x\Vert^2}\text{Tr}\left(P^x_y\circ P^x_{y'}\right)=\frac{1}{2\Vert
x\Vert^2}\text{Tr}\left(\Vert x\Vert^2\cdot\mid\!y\ra\la
y'\!\!\mid+\la
y,y'\ra_\cH\cdot\!\mid\!x\ra\la x\!\!\mid\right)\\
&=&\frac{1}{2\Vert x\Vert^2}\left(\Vert x\Vert^2\left(\la
y',y\ra_\cH+\la y,y'\ra_\cH\right)\right)=\Re(\la
y,y'\ra_\cH)=g(y,y').
\eeas
To prove (\ref{b}), we use (\ref{sym}):
\beas
\zh^r_\zx(P^x_y,P^x_{y'})&=&-\la\zx,[A_y,A_{y'}]\ra_{u*^(\cH)}=
-\frac{1}{2}\text{Tr}\left(\zx\circ[A_y,A_{y'}]\right)\\
&=&-\frac{1}{2}\left\la x,[A_y,A_{y'}]x\right\ra_\cH=
-\frac{1}{2i} \left(\la
A_yx,A_{y'}x\ra_\cH-\la A_{y'}x,A_yx\ra_\cH\right)\\
&=&-\Im(\la iy,iy'\ra_\cH)=\zw(y,y').
\eeas
Finally, (\ref{c}) follows directly from $\zw(y',Jy)=g(y',y)$ and
(\ref{x}).
\ep

The above theorem says that the K\"ahler manifold
$(\cD^1_r(\cH),^r\!\cJ,\zs^r,\zh^r)$ comes from a sort of a
`K\"ahler reduction' of the original linear K\"ahler manifold
$(\cH_\R,J,g,\zw)$. In particular, the symplectic manifold
$\cD^1_r(\cH)$ is the symplectic reduction of $(\cH_\R,\zw)$ with
respect to the isotropic submanifold $S_r=\{ x\in\cH:\Vert
x\Vert=r\}$. The characteristic foliation of $\zw_{\mid_{S_r}}$
consists of orbits of the group $S^1=\{ z\in\C:\vert z\vert=1\}$
acting on $\cH$ by multiplication. The fundamental vector field of
this action is
$$-\widetilde{iI}=p_k\pa_{q_k}-q_k\pa_{p_k}$$
which is simultaneously a Killing vector field for the Riemannian
metric $g$. Therefore $g$ induces a Riemannian metric on
$\cD^1_r(\cH)$, etc.

\section{Smooth manifold structure on $\cP^k(\cH)$}
Recall that the space of non-negatively defined operators from
$gl(\cH)$, i.e. of those $\zr\in gl(\cH)$ which can be written in
the form $\zr=T^\dag T$ for a certain $T\in gl(\cH)$, we denote by
$\cP(\cH)$. It is a cone as being invariant with respect to the
homoteties by $\zl$ with $\zl\ge 0$. The set of density states
$\cD(\cH)$ is distinguished in the cone $\cP(\cH)$ by the equation
$\text{Tr}(\zr)=1$, so we will regard $\cP(\cH)$ and $\cD(\cH)$ as
embedded in $u^*(\cH)$.

The space $\cD(\cH)$ is a convex set in the affine hyperplane in
$u^*(\cH)$, determined by the equation $\text{Tr}(\zt)=1$. The
tangent spaces to this affine hyperplane are therefore canonically
identified with the space of Hermitian operators with trace $0$.
It is known that the set of extreme points of $\cD(\cH)$ coincides
with the set $\cD^1(\cH)$ of pure states, i.e. the set of
one-dimensional orthogonal projectors $\mid x\ra\la x\mid$ (see
Corollary \ref{extreme}). Hence every element of $\cD(\cH)$ is a
convex combination of points from $\cD^1(\cH)$. The space
$\cD^1(\cH)$ of pure states can be identified with the complex
projective space $P\cH\simeq\C P^{n-1}$ via the projection
$\cH\setminus\{ 0\}\ni x\mapsto \mid\! x\ra\la
x\!\!\mid\in\cD^1(\cH)$ which identifies the points of the orbits
of the $\C\setminus\{ 0\}$-group action by complex homoteties. We
have already seen that $\cD^1(\cH)$ is canonically a K\"ahler
manifold. This will be the starting point for the study of
geometry of the set $\cD(\cH)$ of all density states.

\smallskip\noindent
The (co)adjoint action of the group $U(\cH)$ in $u^*(\cH)$ induces
its action on the positive cone $\cP(\cH)$ and on the space of
density states. This action is transitive on pure states but it is
no longer transitive on subsets $\cD^k(\cH)$, $k>1$, where
$\cD^k(\cH)=\cD(\cH)\cap\cP^k(\cH)$ and $\cP^k(\cH)$ consists of
non-negative operators of rank $k$. The rank is understood clearly
as the rank of the corresponding operator (or matrix, if a basis
in $\cH$ is chosen). The intersection of $\cD(\cH)$ with any Weyl
chamber in a Cartan subalgebra in $u^*(\cH)$ is an
$(n-1)$-dimensional simplex, while the intersection of
$\cD^k(\cH)$ is the $(k-1)$-skeleton of this simplex. However, the
dimension of the orbit may vary even for points from a chosen
$\cD^k(\cH)$ if $k>1$. Thus, the set of density states is a union
of smooth manifolds -- orbits of $U(\cH)$ -- but the
differentiable structure of the stratum $\cD^k(\cH)$ is {\it a
priori} not clear (for $k>1$), since the decomposition into orbits
is not a regular foliation, i.e. $\cD^k(\cH)$ is the union of a
family of various submanifolds of $u^*(\cH)$ which differ even by
dimensions. By the differential structure we mean here the
differential structure inherited from $u^*(\cH)$, so that the
smooth curves in $\cD(\cH)$ and hence the tangent spaces are
uniquely defined.

\smallskip\noindent
Our aim in this section is to understand this differential
structure. Of course, the interior of $\cD(\cH)$, namely
$\cD^{n}(\cH)$, is an open subset, so a submanifold, in the affine
subspace of trace=1 Hermitian operators and the real question is
only the boundary, consisting of those density states $\zr$ for
which $\text{det}(\zr)=0$. The best situation would be if the
boundary were a submanifold, but this is not true in dimensions
$n>2$ as we will show later. The stratification into
$U(\cH)$-orbits is too small, since, as it will appear later, the
subsets $\cD^k(\cH)$ are coarser submanifolds in $u^*(\cH)$. We
will show also that the stratification by rank is the maximal one
in the sense that the vectors tangent to $\cD(\cH)$ at
$\zr\in\cD^k(\cH)$ must be tangent to $\cD^k(\cH)$ itself, so the
largest $u^*(\cH)$-submanifold through $\zr\in\cD^k(\cH)$
contained in $\cD(\cH)$ is $\cD^k(\cH)$.

We start with fixing an orthonormal basis in $\cH$ which allows us
to identify $u^*(\cH)$ with the space $u^*(n)$ of Hermitian $n\ti
n$-matrices which is canonically an $n^2$-dimensional real
manifold with respect to the identification
$$u^*(n)\ni (a_{ij})\mapsto
((a_{ii})_{1}^{n},(a_{ij})_{i<j})\in\R^n\ti\C^{n(n-1)/2}.$$ By $\cP(n)$ we
denote the space of non-negatively defined matrices from $u^*(n)$, by
$\cP^k(n)$ the subset of rank $k$ matrices from $\cP(n)$, etc. Let us denote
by $\cP^k_J(n)$ the set of matrices $A=(a_{ij})_{i,j=1}^n\in \cP(n)$ being of
rank $k$ and such that the minor $\text{det}[(a_{rs})_{r,s\in J}]$ associated
with a set of indices $J=\{ i_1,\dots,i_k\}\subset\{ 1,\dots,n\}$ is
non-vanishing.\footnote{The set $J$ is not to be confused in the following
with the complex structure denoted accidentally by the same letter. From the
context, however, the notion of $J$ is always obvious.} The next lemma shows
that any matrix from $\cP^k_J(n)$ can be reconstructed from its rows (or
columns, since it is Hermitian) indexed by $J$.
\begin{lem}\label{l2} Let $A=(a_{ij})_{i,j=1}^n\in \cP^k_J(n)$, so
that the matrix $(a_{rs})_{r,s\in J}$ has the inverse
$(a^{rs})_{r,s\in J}$.  Then the matrix $A$ is uniquely determined
by $\{(a_{ij}):i\in J,j=1,\dots,n\}$ according to the formula
\be\label{1} a_{ij}=\sum_{r,s\in J}{a_{ir}}a^{rs}\overline{a_{js}}.\ee
\end{lem}
{\it Proof.-} The matrix $A$ being non-negatively defined is of
the form $T^\dag T$ for certain $n\ti n$-matrix $T$, so that
$a_{ij}$ is the Hermitian product $\la\za_i,\za_j\ra$ of columns
of $T$ with respect to the standard Hermitian product (\ref{her}).
The matrix $T$ is not uniquely determined. However, the fact that
$A$ is of rank $k$ with the non-vanishing minor associated with
$J$ means that the columns $\za_j,j\in J$ are linearly independent
and span the rest of the columns of $T$. But the Hermitian product
on the subspace in $\C^n$ spanned by $\{\za_j:j\in J\}$ is given
by the formula
\be\label{2} \la x,y\ra_{\C^n}=\sum_{r,s\in J}\la
x,\za_r\ra_{\C^n}\za^{rs}\la\za_s,y\ra_{\C^n},\ee where
$(\za^{rs})_{r,s\in J}$ is the inverse of the matrix
$(\la\za_r,\za_s\ra_{\C^n})_{r,s\in J}$. The proof of (\ref{2}) is
immediate, since the r.h.s. of (\ref{2}) is $\C$-linear with
respect to $y$, anti-linear with respect to $x$ and equals
$\la\za_i,\za_j\ra_{\C^n}$ for $x=\za_i$, $y=\za_j$, $i,j\in J$,
by definition. Since $a_{ij}=\la\za_i,\za_j\ra_{\C^n}$, we get the
formula (\ref{1}) directly from (\ref{2}). \ep

\medskip\noindent
{\bf Remark.} It is worth noticing that the formula (\ref{2}) is
similar to the one describing the Dirac bracket on constraint
manifolds induced by second class constraints.

\bigskip%\noindent
For $J$ as above define a linear map
$$\zF_J:u^*(n)\raa
u^*(k)\ti\C^{(n-k)k}\simeq\R^k\ti\C^{(2nk-k^2-k)/2}\simeq\R^{2nk-k^2}$$
by
\be\label{2a}\zF_J((a_{ij})_{i,j=1}^n)=((a_{ij})_{i,j\in
J},(a_{rs})_{r\notin J,s\in J})).\ee In particular, if we work
with the principal minor, i.e. $J=\{ 1,\dots,k\}$, then $\zF_J$
associates with a Hermitian matrix its first $k$ columns with
removed, say, upper-triangular part which is irrelevant due to
hermicity or, equivalently, its first $k$ rows with removed
lower-triangular part.

\smallskip\noindent
For $A\in u^*(n)$ by $\zF_{J,A}$ we denote the map
$\zF_{J,A}(X)=\zF_J(X)-\zF_J(A)$:
\be\label{map}\zF_{J,A}((x_{ij})_{i,j=1}^n)=((x_{ij}-a_{ij})_{i,j\in
J},(x_{rs}-a_{rs})_{r\notin J,s\in J})).\ee With some abuse of
notation, its restriction to $\cP^k(n)$ we will denote by the same
symbol. It is clear from the above Lemma that the map $\zF_J$ is
continuous and injective on $\cP^k_J(n)$. Thus, for
$A\in\cP^k_J(n)$, also the map $\zF_{J,A}$ is continuous and
injective on $\cP^k_J(n)$.

\smallskip\noindent
Conversely, every point
$$((y_{ij})_{i,j\in J},(y_{rs})_{r\notin
J,s\in J}))$$ of $u^*(k)\ti\C^{(n-k)k}\simeq\R^{2nk-k^2}$,
sufficiently close to $0$, is the value $\zF_{J,A}(X)$ for a
certain $X\in\cP^k_J(n)$. Indeed, adding a small Hermitian matrix
to $(a_{ij})_{i,j\in J}$ will not change its invertibility. Hence
we have to reconstruct $X$ out of $\zF_J(X)$, i.e out of the
columns (and rows, since $X$ should be Hermitian) with indices
belonging to $J$ and knowing that $(x_{ij})_{i,j\in J}$ has an
inverse, say, $(x^{rs})_{r,s\in J}$. Here $x_{ij}=a_{ij}+y_{ij}$
for $j\in J$. An obvious choice is the formula (\ref{1}), i.e.
$$x_{ij}=\sum_{r,s\in J}{x_{ir}}x^{rs}\overline{x_{js}}.$$
The only thing to be checked is that $X=(x_{ij})_{i,j=1}^n$
defined in this way is non-negatively defined and of rank $k$.
Assume, for simplicity of notation, that $J=\{ 1,\dots,k\}$.
First, we can find vectors $\zb_1,\dots,\zb_k\in\C^k$ such that
\be\label{3}x_{ij}=\la\zb_i,\zb_j\ra_{\C^k}\ee
for $i,j=1,\dots,k$. This can be done up to a unitary
transformation. For example, $\zb_i$ can be columns of the matrix
$\sqrt{(x_{ij})_{i,j=1}^k}$. Then, we find (this time unique)
vectors $\zb_{k+1},\dots,\zb_n\in\C^k$ satisfying the conditions
$x_{ij}=\la\zb_i,\zb_j\ra_{\C^k}$, $i=k+1,\dots,n$, $j=1,\dots,k$.
It is easy to see now that, due to the formula (\ref{2}), we have
(\ref{3}) for all $i,j=1,\dots,n$. This immediately implies that
$X$ is non-negatively defined and of rank $k$. Moreover, since
\be\label{4} x_{ij}=\sum_{r,s\in
J}(a_{ir}+y_{ir})a_y^{rs}(\overline{y_{js}}+\overline{a_{js}}),
\ee where $(a_y^{rs})_{r,s\in J}$ is the inverse of the matrix
$(a_{rs}+y_{rs})_{r,s\in J}$, the matrix elements $x_{ij}$
rationally depend on $y_{ml}$, so that $\zF_{J,A}^{-1}$ is smooth,
thus also regular, as a function from a neighbourhhod of $0$ in
$\R^{2nk-k^2}$ into $u^*(n)$, so $\cP^k(n)$ is a submanifold in
$u^*(n)$. To see the image of the differential of $\zF_{J,A}^{-1}$
at $0$, i.e the tangent space ${\rm T}_A\cP^k(n)$, let us consider
the linear (with respect to $y$) part $(v_{ij})$ of the r.h.s. of
(\ref{4}):
\be\label{5} v_{ij}=\sum_{r,s\in
J}(y_{ir}a^{rs}\overline{a_{js}}-a_{ir}a^{rm}y_{ml}a^{ls}\overline{a_{js}}
+a_{ir}a^{rs}\overline{y_{js}}).\ee To see this better, let us
change the orthogonal basis of $\C^n$ for such that $J=\{
1,\dots,k\}$ and $A$ is diagonal, $a_{ii}=\zl_i$, $\zl_i=0$ for
$i>k$. Then one can easily find that (\ref{5}) takes the form
\be\label{6} v_{ij}=\begin{cases}0,& \text{if \ }i,j>k \\
y_{ij}, & \text{if \ }j\leq k.\end{cases} \ee This means that in
the image are arbitrary Hermitian matrices $V=(v_{ij})_{i,j=1}^n$
such that $v_{ij}=0$ for $i,j>k$, that can be written in a
coordinate-free way as $\la Vx,y\ra_{\C^n}=0$ for all
$x,y\in\text{Ker}(A)$. Note that the manifold $\cP^k(\cH)$ is
connected. Indeed. it consists of connected orbits of the group
$U(\cH)$ which meet a Weyl chamber as the $(k-1)$-dimensional
skeleton of a simplex. However, the connected components of this
skeleton are identified by the action of the Weyl group, so they
form topologically a $(k-1)$-dimensional simplex which is
obviously connected. Therefore we have proved the following.

\begin{theo}\label{t2} Let $A\in\cP^k_J(n)$. Then the map
$\zF_{J,A}:\cP^k(n)\raa\R^{2nk-k^2}$ defined by (\ref{map}) is a
local homeomorphism from a neighbourhood of $A$ in $\cP^k(n)$ onto
a neighbourhood of $0$ in
$u^*(k)\ti\C^{(n-k)k}\simeq\R^{2nk-k^2}$. Moreover, the collection
of the maps $\zF_{J,A}^{-1}:\cW_{J,A}\raa\cP^k(n)\subset u^*(n)$
defined on sufficiently small neighbourhoods $\cW_{J,A}$ of $0$ by
the formula (\ref{4}) constitutes a smooth manifold structure on
$\cP^k(n)$ which makes it into a smooth and connected submanifold
of $u^*(n)$. The tangent space ${\rm T}_A\cP^k(n)$, viewed as a
subspace of $u^*(n)$ consists of matrices $V\in u^*(n)$ satisfying
$\la Vx,y\ra_{\C^n}=0$ for all $x,y\in\text{Ker}(A)$.
\end{theo}
{\bf Remark.} In section 5 we obtain the manifold structure on
$\cP^k(n)$ much simpler as the structure of an $GL(n,\C)$-orbit.
But we find that Lemma \ref{l2} and Theorem \ref{t2} are of some
interest {\it per se} providing explicit coordinate systems.

\bigskip\noindent
The next theorem shows that smooth curves in $u^*(n)$ which lay in
$\cP(n)$ cannot cross $\cP^k(n)$ transversally, i.e. $\cP^k(n)$ is
in a sense an edge for $\cP^{k+1}(n)$ if $k<n-1$.
\begin{theo}\label{t3} Let $\zg:\R\raa u^*(n)$ be a smooth curve in the
space of Hermitian matrices which lies entirely in $\cP(n)$. Then
$\zg$ is tangent to the stratum $\cP^k(n)$ it belongs, i.e.
$\zg(t)\in\cP^k(n)$ implies $\dot{\zg}(t)\in{\rm
T}_{\zg(t)}\cP^k(n)$.
\end{theo}
{\it Proof.-} Of course, it is enough to prove the above for an
arbitrary $t\in\R$, say, $t=0$. Assume therefore that $A=\zg(0)\in
\cP^k(n)$. Take $x\in\text{Ker}(A)$. Since
$$\left\la\frac{\zg(\zD t)-\zg(0)}{\zD t}x,x\right\ra\ge 0$$
for $\zD t\ge 0$, we have $\la\dot{\zg}(0)x,x\ra\ge 0$. Taking in
turn $\zD t\leq 0$ we see in a similar way that
$\la\dot{\zg}(0)x,x\ra\leq 0$, so
\be\label{7}\la\dot{\zg}(0)x,x\ra= 0.\ee
By polarization of (\ref{7}) we get
\be\label{8}\la\dot{\zg}(0)x,y\ra+\la\dot{\zg}(0)y,x\ra= 0\ee for all
$x,y\in\text{Ker}(A)$. But $\dot{\zg}(0)$ is Hermitian, so
$$\la\dot{\zg}(0)y,x\ra=\la y,\dot{\zg}(0)x\ra$$ and (\ref{8})
yields that the real part $\Re(\la\dot{\zg}(0)x,y\ra)$ is $0$ for
all $x,y\in\text{Ker}(A)$. On the other hand, the kernel of $A$ is
a complex subspace and
$$\Re(\la\dot{\zg}(0)x,i\cdot y\ra)=\Im(\la\dot{\zg}(0)x,y\ra)$$
so
\be\label{9}\la\dot{\zg}(0)x,y\ra=0\ee for
all $x,y\in\text{Ker}(A)$. But, according to Theorem \ref{t2},
(\ref{9}) means that $\dot{\zg}(0)\in{\rm T}_A\cP^k(n)$. \ep

\section{Smooth stratification of density states}

The set $\cD(\cH)$ of density states on $\cH$ is the intersection
of the cone $\cP(\cH)$ with the affine subspace $\{ A\in
u^*(\cH):\text{Tr}(A)=1\}$ or, in other words, it is the level set
of the function $\text{Tr}:\cP(\cH)\raa\R$ corresponding to the
value $1$. Since $\text{Tr}(t\zr)=t\text{Tr}(\zr)$ and $\cP^k$ is
invariant with respect to homoteties with positive $t$, it is
clear that $\text{Tr}$ is a regular function on each $\cP^k(\cH)$,
so that $\cD^k(\cH)$ is canonically a smooth manifold. Since
topologically $\cP^k(\cH)\simeq\cD^k(\cH)\ti\R$, the manifolds
$\cD^k(\cH)$ are connected. All these observations together with
Theorems \ref{t2} and \ref{t3} can be summarized in the following.
\begin{theo}\label{t00} The spaces $\cD^k(\cH)$ of density
states of rank $k$, $k=1,\dots,n$, are smooth and connected
submanifolds in $u^*(\cH)$ of (real) dimension $2nk-k^2-1$. The
tangent space ${\rm T}_\zr\cD^k(\cH)$ is characterized as the
space of those Hermitian operators $T$ of trace $0$ which satisfy
$\la Tx,y\ra=0$ for all $x,y\in\text{Ker}(\zr)$. Moreover, the
stratification into submanifolds $\cD^k(\cH)$ is maximal in the
sense that every smooth curve in $u^*(\cH)$, which lies entirely
in $\cD(\cH)$, at every point is tangent to the strata
$\cD^k(\cH)$ to which it actually belongs.
\end{theo}
\begin{cor} The boundary $\partial\cD(\cH)=\bigcup_{k<n}\cD^k(\cH)$
of the set of density states is not a smooth submanifold of
$u^*(\cH)$ if $n=\text{dim}\cH>2$.
\end{cor}
{\it Proof.-} If $n>2$ then the boundary $\partial\cD(\cH)$ has at
least two different strata and the vectors orthogonal to, say, the
stratum $\cD^1(\cH)$ of pure states are not tangent to
$\partial\cD(\cH)$. But the dimension of $\cD^1(\cH)$ is smaller
than the topological dimension of $\partial\cD(\cH)$.  \ep

\medskip\noindent {\bf Remark.} It is well known that for $n=2$ the
convex set of density states is affinely equivalent to the
three-dimensional ball and its boundary -- to the two-dimensional
sphere, so it is a smooth manifold.

\medskip\noindent
The last problem concerning the geometry of density states we will consider is
the question of affine parts of the manifolds $\cD^k(\cH)$. It is motivated by
the fact that the set $\cD^1(\cH)$ of pure states is exactly the set of
extremal elements of $\cD(\cH)$, so it does not contain intervals, but the
other strata $\cD^k(\cH)$ with $k>1$ must do as shows the following theorem.
Recall that a non-empty closed convex subset $K_0$ of a closed convex set $K$
is called a {\it face} (or extremal subset) of $K$ if any closed segment in $K$
with an interior point in $K_0$ lies entirely in $K_0$; a point $x$ is called
an {\it extreme} point of $K$ if the set $\{ x\}$ is a face of $K$.
\begin{theo} If $\zr\in\cD^k(\cH)$ then the affine space in
$u^*(\cH)$ which is tangent to $\cD^k(\cH)$ at $\zr$ intersects
$\cD(\cH)$ along a $(k^2-1)$-dimensional convex body which is
affinely equivalent to the set $\cD(k)$ of density states in
dimension $k$. This convex body is exactly the face of $\cD(\cH)$
at $\zr$. In other words, the face of $\cD(\cH)$ at
$\zr\in\cD^k(\cH)$ is affinely equivalent to $\cD(k)$.
\end{theo}
{\it Proof.-} Let us take coordinates in $u^*(\cH)$, i.e. let us
chose an orthonormal basis in $\cH$, in which $\zr$ is represented
by a diagonal matrix $(\zr_{ij})$, $\zr_{ij}=\zd^i_j\zl_i$, where
$\zl_i=0$ for $i>k$. According to the form of ${\rm
T}_\zr\cD^k(\cH)$, matrices $(x_{ij})$ which belong to $\zr+{\rm
T}_\zr\cD^k(\cH)$ have entries $x_{ij}$ with $i,j>k$ equal to $0$.
If they belong as well to $\cD(\cH)$, also $x_{ij}=0$ if $i>k$ or
$j>k$. Indeed, since $x_{ij}=\la\za_i,\za_j\ra$ for certain
vectors $z_i\in\C^n$, we have $x_{ii}=\|\za_i\|^2=0$, so
$\za_i=0$, for $i>k$, and further $x_{ij}=\la\za_i,\za_j\ra=0$ if
$i>k$ or $j>k$. In other words, the only non-zero part of $X$ is
the block $(x_{ij})_{i,j=1}^k$ which is therefore an element of
$\cD(k)$. Conversely, every matrix $X$ with such a block form
belongs simultaneously to $\cD(\cH)$ and, since $(X-\zr)_{ij}=0$
for $i,j>k$, to $\zr+{\rm T}_\zr\cD^k(\cH)$. To see that
$(\zr+{\rm T}_\zr\cD^k(\cH))\cap\cD(\cH)$ is exactly the face of
$\cD(\cH)$ at $\zr$, consider a segment in $\cD(\cH)$ for which
$\zr$ is an interior point. The open segment is clearly a smooth
curve in $\cD(\cH)$, so, in view of Theorem \ref{t00}, it is
tangent to $\cD^k(\cH)$ at $\zr$, thus belongs entirely to
$\zr+{\rm T}_\zr\cD^k(\cH)$. \ep
\begin{cor}\label{extreme}
Extremal points of $\cD(\cH)$ are exactly pure states.
\end{cor}

\section{Geometry of $u^*(\cH)$}

Let us mention that a major part of what has been said about the
differential structure of the space $\cP^k(\cH)$ of rank-$k$
positive operators can be repeated for the space of all rank-$k$
Hermitian operators. Denote by $u^*_{k_+,k_-}(\cH)$ the set of
those Hermitian operators $\zx$ whose spectrum contains $k_+$
positive and $k_-$ negative eigenvalues (counted with
multiplicities), respectively. Thus the rank of $\zx$ is
$k=k_++k_-$ and $\cP^k(n)=u^*_{k,0}(n)$.

Fixing an orthogonal basis in $\cH$ will identify
$u^*_{k_+,k_-}(\cH)$ with the space $u^*_{k_+,k_-}(n)$ of $n\ti n$
Hermitian matrices of rank $k$ with the corresponding spectrum.
Denote by $D^{k_+}_{k_-}$ the diagonal matrix
$diag(1,\dots,1,-1,\dots,-1,0,\dots,0)$ with $1$ coming
$k_+$-times and $-1$ coming $k_-$-times. Denote by
$\la\cdot,\cdot\ra_{k_+,k_-}$ the `semiHermitian' product in
$\C^n$ represented by $D^{k_+}_{k_-}$:
\be\la a,b\ra_{k_+,k_-}=\sum_{j=1}^{k_+}\overline{a_j}b_j-
\sum_{j=k_++1}^{k_++k_-}\overline{a_j}b_j.
\ee
It is easy to see the following.
\begin{prop}\label{pr1} Any $\zx=(a_{ij})\in u^*_{k_+,k_-}(n)$ can
be written in the form $\zx=T^\dag D^{k_+}_{k_-}T$ for certain
$T\in GL(n,\C)$. In other words the entries of the matrix $\zx$
are semiHermitian products $a_{ij}=\la\za_i,\za_j\ra_{k_+,k_-}$,
where $\za_i$ denotes the $i$th column of $T$.
\end{prop}
{\it Proof.} We can diagonalize $\zx$ by means of an unitary
matrix $U$,
$$U \zx U^\dag=diag(\zl_1,\dots,\zl_n),$$
where $\zl_1\ge\dots\ge\zl_n$, so $\zl_1,\dots,\zl_{k_+}>0$ and
$\zl_{k_++1},\dots,\zl_{k_++k_-}<0$. Hence $\zx=T^\dag
D^{k_+}_{k_-}T$ for $T=CU$ with
$$C=diag\left({\sqrt{\vert\zl_1\vert}},\dots,
{\sqrt{\vert\zl_{k_++k_-}\vert}},1\dots,1\right).$$ \ep

Now, we can reformulate Lemma \ref{l2} for $u^*_{k_+,k_-}(n)$
instead of $\cP^k(n)$. The proof is essentially the same with the
difference that we use the semiHermitian product
$\la\cdot,\cdot\ra_{k_+,k_-}$ in $\C^n$ instead of
$\la\cdot,\cdot\ra_{\C^n}$.
\begin{lem}\label{l3} Let $\zx=(a_{ij})_{i,j=1}^n\in
u^*_{k_+,k_-}(n)$. Assume that the matrix $(a_{rs})_{r,s\in J}$
has the inverse $(a^{rs})_{r,s\in J}$ for certain
$k=(k_++k_-)$-element subset $J=\{ j_1,\dots,j_k\}\subset\{
1,\dots,n\}$.  Then the matrix $\zx$ is uniquely determined by
$\{(a_{ij}):i\in J,j=1,\dots,n\}$ according to the formula
\be\label{1x} a_{ij}=\sum_{r,s\in J}{a_{ir}}a^{rs}\overline{a_{js}}.\ee
\end{lem}
One can now prove that $u^*_{k_+,k_-}(\cH)$ are submanifolds of
$u^*(\cH)$ in completely parallel way to the case of $\cP^k(\cH)$.
However, Proposition \ref{pr1} suggest an easier (although less
constructive) way to do it. Namely, we can see
$u^*_{k_+,k_-}(\cH)$ as an orbit of a natural $GL(\cH)$ action on
$u^*(\cH)$.
\begin{theo} The family
\be\{ u^*_{k_+,k_-}(\cH):k_+,k_-\ge 0, k=k_++k_-\le n\}
\ee
of subsets of $u^*(\cH)$ is exactly the family of orbits of the
smooth action of the group $GL(\cH)$ given by
\be\label{action} GL(\cH)\ti u^*(\cH)\ni (T,\zx)\mapsto T\zx T^\dag\in u^*(\cH).
\ee
In particular, every $u^*_{k_+,k_-}(\cH)$ is a connected
submanifold of $u^*(\cH)$ and the tangent space to
$u^*_{k_+,k_-}(\cH)$ at $\zx$ is characterized by
\be\label{p} B\in\text{T}_\xi u^*_{k_+,k_-}(\cH)\Leftrightarrow
\forall x,y\in\text{Ker}(\zx )\ [\la Bx,y\ra_\cH=0].
\ee
Moreover, the following are equivalent:
\begin{description}
\item{(1)} $u^*_{k_+,k_-}(\cH)$ intersects $\cP(\cH)$; \item{(2)}
$u^*_{k_+,k_-}(\cH)$ is contained in $\cP(\cH)$; \item{(3)}
$k_-=0$; \item{(4)} $u^*_{k_+,k_-}(\cH)=\cP^{k}(\cH)$,
$k=k_++k_-$.
\end{description}
\end{theo}
{\it Proof.} The proof that (\ref{action}) is a group smooth
action is straightforward. Proposition \ref{pr1} shows that
$u^*_{k_+,k_-}(\cH)$ is contained in the $GL(\cH)$-orbit of
$D^{k_+}_{k_-}$.

On the other hand, although the spectrum is not fixed on every
$GL(\cH)$-orbit, the number of positive and the number of negative
eigenvalues (counted with multiplicities) are fixed along the
orbit. Indeed, if $\la x,\zx x\ra_\cH>0$ (resp. $\la x,\zx
x\ra_\cH<0$) for $x$ in a $k_+$-dimensional (resp.
$k_-$-dimensional) linear subspace $V_+$ (resp. $V_-$), then $\la
x,T\zx T^\dag x\ra_\cH=\la T^\dag x,\zx T^\dag x\ra_{\cH}>0$
(resp. $\la x,T\zx T^\dag x\ra_\cH=\la T^\dag x,\zx T^\dag
x\ra_{\cH}<0$) for $x$ in a $k_+$-dimensional (resp.
$k_-$-dimensional) linear subspace $(T^\dag)^{-1}(V_+)$ (resp.
$(T^\dag)^{-1}(V_-)$).

The corresponding infinitesimal action of $v\in gl(\cH)$ is $\zx
\mapsto v\zx +\zx v^\dag$ and the operators $\xi_v=v\zx +\zx v^\dag$
clearly satisfy $\la x,\zx_v y\ra_\cH=0$ for all
$x,y\in\text{Ker}(\zx )$. Conversely, if for certain $B\in
u^*(\cH)$ we have $\la Bx,y\ra_\cH=0$ for all
$x,y\in\text{Ker}(\zx )$, then $B$ can be written in the form
$v\zx +\zx v^\dag$. To see this, consider the splitting
$\cH=V_1\oplus V_2$, where $V_2=\text{Ker}(\zx )$ and
$V_1=V_2^\bot$. According to this splitting $\zx $ can be written
in the operator matrix form
$$\zx =\left(\begin{array}{cc}\zx_1&0\\ 0&0\end{array}\right),$$
where $\zx_1$ is Hermitian and invertible. Similarly, $B$ has the
form
$$B=\left(\begin{array}{cc}B_{11}&B_{12}\\ B_{21}&0\end{array}\right),$$
where $B_{11}^\dag=B_{11}$ and, in the obvious sense,
$B_{21}=B_{12}^\dag$. Now, it is easy to see that $B=v\zx +\zx
v^\dag$, where
$$v=\left(\begin{array}{cc}\frac{1}{2}B_{11}\zx_1^{-1}&\zx_1B_{12}\\
B_{21}\zx_1^{-1}&0\end{array}\right)$$ that proves (\ref{p}).

Finally, if $u^*_{k_+,k_-}(\cH)$ intersects $\cP(\cH)$, then it
contains an element with non-negative spectrum. But the signs of
the elements of the spectrum are constant along a $GL(\cH)$-orbit
which means that $k_-=0$ and
$u^*_{k_+,k_-}(\cH)=\cP^k(\cH)\subset\cP(\cH)$. \ep

\noindent Note that the fundamental vector fields
$\wt{a}(\zx)=-a\zx-\zx a^\dag$ of the $GL(\cH)$-action satisfy the
commutation rules $[\wt{a},\wt{b}]_{vf}=\wt{[a,b]_-}$

The next results shows that the foliation into submanifolds
$u^*_{k_+,k_-}(\cH)$ can be obtained directly from tensors $\zL$
and $R$. We know already that the (generalized) distribution
$D_\zL$ induced by $\zL$ is generated by vector fields
$\zL_A(\zx)=\#\zL_\zx(\wh{A})=[A,\zx]$ and the (generalized
distribution $D_R$ induced by $R$ is generated by vector fields
$R_A(\zx)=\# R_\zx(\wh{A})=[A,\zx]_+$. The following is
straightforward.
\begin{theo} The family $\{\zL_A,R_A:A\in u^*(\cH)\}$ of linear vector
fields on $u^*(\cH)$ is the family of fundamental vector fields of
the $GL(\cH)$-action:
\bea \zL_A(\zx)&=&\frac{1}{i}(A\zx-\zx
A)=-(iA)\zx-\zx(iA)^\dag=\wt{iA}(\zx),\\
R_A(\zx)&=& A\zx+\zx A=A\zx+\zx A^\dag=-\wt{A}(\zx).
\eea
In particular,
\be [\zL_A,\zL_B]_{vf}=\zL_{[A,B]},\quad
[ R_A,R_B ]_{vf}=\zL_{[A,B]},\quad [R_A,\zL_B]_{vf}= R_{[A,B]},
\ee
so the (generalized) distribution induced by jointly by the
tensors $\zL$ and $R$ is completely integrable and
$u^*_{k_+,k_-}(\cH)$ are the maximal integrate submanifolds.
\end{theo}
\begin{cor} The generalized distributions $D_{gl}=D_R+D_\zL$,
$D_\zL$ and $D_0=D_R\bigcap D_\zL$ on $u^*(\cH)$ are involutive
and can be integrated to generalized foliations $\cF_{gl}$,
$\cF_\zL$, and $\cF_0$, respectively. The leaves of the foliation
$\cF_{gl}$ are the orbits of the $GL(\cH)$ action $\zx\mapsto T\zx
T^\dag$, the leaves of $\cF_\zL$ are the orbits of the
$U(\cH)$-action.
\end{cor}

Denote by $\wt{\cJ}$ and $\wt{\cR}$ the $(1,1)$-tensors on
$u^*(\cH)$, viewed as a vector bundle morphism induced by the
contravariant tensors $\zL$ and $R$, respectively,
\beas\wt{\cJ},\wt{R}&:&\text{T}u^*(\cH)\raa \text{T}u^*(\cH),\\
\wt{\cJ}_\zx(A)&=&[A,\zx]=\zL_\zx(A),\\
\wt{\cR}_\zx(A)&=&[A,\zx]_+=R_\zx(A),
\eeas
where $A\in u^*(\cH)\simeq\text{T}_\zx u^*(\cH)$. The image of
$\wt{\cJ}$ is $D_\zL$ and the image of $\wt{\cR}$ is $D_R$.
\begin{lem} The tensors $\wt{\cJ}$ and $\wt{\cR}$ commute and
\be\wt{\cJ}_\zx\circ\wt{\cR}_\zx(A)=
\wt{\cR}_\zx\circ\wt{\cJ}_\zx(A)=[A,\zx^2].
\ee
\end{lem}
{\it Proof.} We have
$$\wt{\cJ}_\zx\circ{\cR}_\zx(A)=[[A,\zx]_+,\zx].$$
But, as easily seen,
\be\label{RJ}[[A,\zx]_+,\zx]=[A,\zx^2]=[[A,\zx],\zx]_+=
\wt{\cR}_\zx\circ\wt{\cJ}_\zx(A).
\ee
\ep

Recall that $U(\cH)$-orbits ${\cO}$, i.e. the orbits with respect
to the action of the subgroup $U(\cH)\subset GL(\cH)$, carry
canonical symplectic structures $\zh^{{\cO}}$. The symplectic
structures $\zh^{{\cO}}$ is $U(\cH)$-invariant, i.e.
$({\cO},\zh^{{\cO}})$ is a homogeneous symplectic manifold. We
will show that this symplectic structure is a part of a canonical
K\"ahler structure. We know already this structure for the orbits
$\cP^1_r(\cH)$.

Recall also that on $u^*(\cH)$ we have the Riemannian metric
induced by the scalar product $\la
A,B\ra_{u^*}=\frac{1}{2}\text{Tr}(AB)$ on $u^*(\cH)$.
\begin{theo}\begin{description} \item{(a)} The image of
$\wt{\cJ}_\zx$ is $\text{T}_\zx{\cO}$ and
$\text{Ker}(\wt{\cJ}_\zx)$ is the orthogonal complement of
$\text{T}_\zx{\cO}$.

\item{(b)} $\wt{\cJ}_\zx^2$ is a self-adjoint (with respect to
$\la\cdot,\cdot\ra_{u^*}$) and negatively defined operator on
$\text{T}_\zx{\cO}$.

\item{(c)} The $(1,1)$-tensor $\cJ$ on $u^*(\cH)$ defined by
\be\label{J}
\cJ_\zx(A)=\left(-(\wt{\cJ}_\zx)^2_{\mid\text{T}_{\zx}{\cO}}\right)^{-\frac{1}{2}}
\wt{\cJ}_\zx(A)
\ee
induces an $U(\cH)$-invariant complex structure $\cJ$ on every
orbit ${\cO}$.

\item{(d)} The tensor
\be\zg^{{\cO}}_\zx(A,B)=\zh^{{\cO}}_\zx(A,\cJ_\zx(B))
\ee
is an $U(\cH)$-invariant Riemannian metric on ${\cO}$ and
\be\label{46}\zg^{{\cO}}_\zx(\cJ_\zx(A),B)=\zh^{{\cO}}_\zx(A,B).
\ee
In particular, $({\cO},\cJ,\zh^\cO,\zg^\cO)$ is a homogeneous
K\"ahler manifold. Moreover, if $\zx\in u^*(\cH)$ is a projector
and $\zx\in\cO$, then $\cJ_\zx=\wt{\cJ}_\zx$ and $\zg^\cO(A,B)=\la
A,B\ra_{u^*}$.
\end{description}
\end{theo}
{\bf Remark.} The tensor $\cJ$ is canonically and globally
defined. It is however not smooth as a tensor field on $u^*(\cH)$.
It is smooth on the open-dense subset of regular elements and, of
course, on every $U(\cH)$-orbit separately.

\medskip\noindent
{\it Proof.} \begin{description} \item{(a)} The vector fields
$\zL_A(\zx)=[A,\zx]=\wt{\cJ}_\zx(A)$ are fundamental vector fields
of the $U(\cH)$-action, so $\text{T}_\zx{\cO}$ is the image of
$\wt{\cJ}_\zx$. Moreover, the invariance of the Riemannian metric
$\la A,B\ra_{u^*}$,
\be\label{proof1}\la\wt{\cJ}_\zx(A),B\ra_{u^*}=\la[A,\zx],B\ra_{u^*}= -\la
A,\wt{\cJ}_\zx(B)\ra_{u^*},
\ee
implies that
$$B\in\text{Ker}(\wt{\cJ}_\zx)\ \Leftrightarrow
B\bot\wt{\cJ}_\zx(u^*(\cH)).
$$

\item{(b)} The identity (\ref{proof1}) means that
$\wt{\cJ}_\zx^\ti=-\wt{\cJ}_\zx$, where $\wt{\cJ}_\zx^\ti$ is the
adjoint operator to $\wt{\cJ}_\zx$ with respect to the scalar
product $\la A,B\ra_{u^*}$. Consequently,
\be\label{proof2}(\wt{\cJ}_\zx^2)^\ti=\wt{\cJ}_\zx^2.
\ee
Moreover, $\wt{\cJ}_\zx^2$ is negatively defined on
$\text{T}_\zx{\cO}$, since
$$\la\wt{\cJ}_\zx^2(A),A\ra_{u^*}=\la[[A,\zx],\zx],A\ra_{u^*}=
-\la[A,\zx],[A,\zx]\ra_{u^*}<0,
$$
for $[A,\zx]\in\text{T}_\zx{\cO}$, $[A,\zx]\ne 0$.

\item{(c)} The tensor $\wt{\cJ}$ is clearly $U(\cH)$-invariant:
\be\label{proof3}
\wt{\cJ}_{U\zx U^\dag}(UAU^\dag)=[UAU^\dag,U\zx U^\dag]=
U[A,\zx]U^\dag=U(\wt{\cJ}_\zx(A))U^\dag,
\ee
so $U({\cal H})$-invariant is the tensor $\left(-\wt{\cJ}^2\right)^{-\frac{1}{2}}$ and
its composition $\cJ$. The tensor $\cJ$ defines an almost
complex structure on every orbit $\cO$, since
$$\left[\left(-\wt{\cJ}^2\right)^{-\frac{1}{2}}\wt{\cJ}\right]^2=
\left(-\wt{\cJ}^2\right)^{-1}\wt{\cJ}^2=-I.
$$
To show that this almost complex structure is integrable, it is
sufficient to show that the distribution $\cN$ in the complexified
tangent bundle $\text{T}\cO\ot\C$ which consists of
$i$-eigenvectors of (complexified) $\cJ$ is involutive. Since
$\cJ$, and therefore $\cN$, is invariant, it is sufficient to check it at one
point, say $\zx\in\cO$ with respect to the complexified Lie
algebra $gl(\cH)=u^*(\cH)\ot\C$ equipped with the bracket
$[a,b]=\frac{1}{i}[ab-ba]$.

Let $-\zk_1^2,\dots,-\zk_m^2$, where $\zk_1,\dots,\zk_m>0$, be the
eigenvalues of $(\wt{\cJ}_\zx^2)_{\mid\text{T}_{\zx}{\cO}}$
counted with multiplicities. The complexified $\wt{\cJ}_\zx$,
which with some abuse of notation we will denote by the same
symbol, has therefore eigenvalues $\pm i\zk_k$ with eigenvectors
$a_k^\pm$, $k=1,\dots,m$ and $\cJ_\zx(a_k^\pm)=\pm ia_k^\pm$. Thus
$\cN_\zx$ is spanned by the vectors $a^+_k, k=1,\dots,m$, i.e.
eigenvectors of $\wt{\cJ}_\zx$, $\wt{\cJ}_\zx(a^+_k)=i\zk_ka^+_k$
with positive $\zk_k$. This space is clearly a Lie subalgebra in
$gl(\cH)$, since
\beas\cJ_\zx([a^+_k,a^+_l])&=&[[a^+_k,a^+_l],\zx]=
[[a^+_k,\zx],a^+_l]+[a^+_k,[a^+_l,\zx]]\\
&=&[i\zk_ka^+_k,a^+_l]+[a^+_k,i\zk_la^+_l]=i(\zk_k+\zk_l)[a^+_k,a^+_l],
\eeas
the vector $[a^+_k,a^+_l]$, if non-zero, is again an eigenvector
of $\wt{\cJ}_\zx$ corresponding to a `positive' eigenvalue
$i(\zk_k+\zk_l)$.

\item{(d)} The tensor
$$\zg^{{\cO}}_\zx(A,B)=\zh^{{\cO}}_\zx(A,\cJ_\zx(B))
$$
is clearly $U(\cH)$-invariant. From (\ref{proof1}) and
(\ref{proof2}) it follows that $\cJ_\zx^\ti=-\cJ_\zx$. Since
$\wt{\cJ}$ and $\cJ$ clearly commute,
$\cJ_\zx([A,\zx])=[\cJ_\zx(A),\zx]$, in view of (\ref{sym}),
\bea\label{skew}\zh^{{\cO}}_\zx([A,\zx],\cJ_\zx([B,\zx]))&=&
\la[A,\zx],\cJ_\zx(B)\ra_{u^*(\cH)}=
\la-\cJ_\zx([A,\zx]),B\ra_{u^*(\cH)}\\
&=&-\zh^{{\cO}}_\zx(\cJ_\zx([A,\zx]),[B,\zx]).\nonumber
\eea
This immediately implies that $\zg^\cO$ is symmetric and proves
(\ref{46}). But (\ref{sym}) implies also that
\bea\label{ko}\zg^\cO_\zx([A,\zx],[A,\zx])&=&\zh^{{\cO}}_\zx([A,\zx],\cJ_\zx([A,\zx]))=
\la[A,\zx],\cJ_\zx(A)\ra_{u^*(\cH)}\\
&=& \la A,-\wt{\cJ}_\zx\cJ_\zx(A)\ra_{u^*(\cH)}.\nonumber
\eea
But
$$-\wt{\cJ}_\zx\cJ_\zx=\left(-\wt{\cJ}^2\right)^{\frac{1}{2}}
$$
is a positive operator, so
$$\zg^\cO_\zx([A,\zx],[A,\zx])>0$$ for $[A,\zx]\ne 0$.

Finally, if $\zx$ is a projector, $\zx^2=\zx$, then  (cf.
(\ref{u}))
$$\wt{\cJ}_\zx^2([A,\zx])=-[A,\zx],$$
so $\cJ_\zx=\wt{\cJ}_\zx$ and (cf. (\ref{ko}))
$$\zg^\cO_\zx([A,\zx],[B,\zx])=
\la[A,\zx],\cJ_\zx(B)\ra_{u^*(\cH)}=
\la[A,\zx],[B,\zx]\ra_{u^*(\cH)}.
$$
\end{description}
\ep

We have some similar results for the tensor $\wt{\cR}$ which
however are not completely analogous, since the distribution $D_R$
is not globally integrable. The proofs are analogous, so we omit
them.
\begin{theo}\begin{description}
\item{(a)} The image $D_R(\zx)$ of $\ \wt{\cR}_\zx$ is the
orthogonal complement of $\text{Ker}(\wt{\cR}_\zx)$.

\item{(b)} $\wt{\cR}_\zx^2$ is a self-adjoint (with respect to
$\la\cdot,\cdot\ra_{u^*}$) and positively defined operator on
$D_R(\zx)$.

\item{(c)} The $(1,1)$-tensor $\cR$ on $u^*(\cH)$ defined by
\be\label{J1}
\cR_\zx(A)=\vert(\wt{\cR}_\zx)_{\mid D_R(\zx)}\vert^{-1}\circ
\wt{\cR}_\zx(A)
\ee
satisfies $\cR^3=\cR$.
\end{description}
\end{theo}
\begin{cor}The distribution $D_0$ is the image of
${\cJ}_\zx\circ\cR_\zx=\cR_\zx\circ\cJ_\zx$. In other words,
$D_0(\zx)=\{[A,\zx^2]:A\in u^*(\cH)\}$. Moreover, the foliation
$\cF_0$ is $U(\cH)$-invariant, $\cJ$-invariant and
$\cR$-invariant, so that $\cJ$ and $\cR$ induce on leaves of
$\cF_0$ a complex and a product structure, respectively. The
leaves of the foliation $\cF_0$ are also canonically symplectic
manifolds with symplectic structures being restrictions of
symplectic structures on the leaves of $\cF_\zL$, so the leaves of
$\cF_0$ are K\"ahler submanifolds of the $U(\cH)$-orbits in
$u^*(\cH)$.
\end{cor}
{\it Proof.} The image of
$\cJ_\zx\circ\cR_\zx=\cR_\zx\circ\cJ_\zx$ is clearly contained in
$D_0$. Conversely, let $B\in D_0(\zx)=D_\zL\bigcap D_\R$.
According to (\ref{RJ}), $D_0(\zx)$ is invariant with respect to
both: $\cJ_\zx$ and $\cR_\zx$ and $\cJ_\zx$ and $\cR_\zx$ are
injective, thus surjective, on $D_0(\zx)$. The distribution $D_0$
is therefore generated by vector fields $X_A(\zx)=[A,\zx^2]$. It
is a matter of simple calculations to show that these vector
fields commute with the fundamental vector fields $\zL_B$ of the
$U(\cH)$ as $[X_A,\zL_B]_{vf}=X_{[B,A]}$ that shows $U(\cH)$
invariance of $D_0$. One can also easily seen that the
restrictions of $\zx^\cO$ to the leaves of $\cF_0$ contained in
$\cO$ are non-degenerate. It follows also directly from the
explicit calculations we present below. \ep

Let us explain the above theorem in local coordinates, i.e. for
the case of matrices. Suppose that $\zx=diag(\zl_1,\dots\zl_n)\in
u^*(n)$ is a diagonal matrix. For simplicity, it is better to
start already with the complexified structures, i.e. with
$gl(n)=u^*(n)\ot\C$ equipped with the bracket
$[a,b]=\frac{1}{i}(ab-ba)$ and the Hermitian product $\la
a,b\ra_{gl}=\frac{1}{2}\text{Tr}(a^\dag b)$, so that $u^*(n)$ is a
real Lie subalgebra in $gl(n)$ with the induced scalar product.
Let $E^k_l$ be the matrix whose the only non-zero entry is $1$ at
$k$th row and $l$th column. We have
\be
\la E^k_l,E^r_s\ra_{gl}=\frac{1}{2}(\zd_k^r\zd_s^l),
\ee
\be
[E^k_l,E^r_s]=-i(\zd_l^rE^k_s-\zd_s^kE^r_l),
\ee
and
\be
[E^k_l,E^r_s]_+=\zd_l^rE^k_s+\zd_s^kE^r_l,
\ee
so that
\be \wt{\cJ}_\zx(E^k_l)=[E^k_l,\zx]=i(\zl_k-\zl_l)E^k_l.
\ee
and \be \wt{\cR}_\zx(E^k_l)=[E^k_l,\zx]_+=(\zl_k+\zl_l)E^k_l.
\ee
In particular,
\be\wt{\cJ}_\zx\circ\wt{\cR}_\zx(E^k_l)=[E^k_l,\zx^2]=i(\zl_k^2-\zl_l^2)E^k_l.
\ee
Consequently, \be \wt{\cJ}^2_\zx(E^k_l)=-(\zl_k-\zl_l)^2E^k_l
\ee
and
\be \wt{\cR}^2_\zx(E^k_l)=(\zl_k+\zl_l)^2E^k_l,
\ee
so that \be \cJ_\zx(E^k_l)=i\cdot sgn(\zl_k-\zl_l)E^k_l.
\ee
and
\be
\cR_\zx(E^k_l)=sgn(\zl_k+\zl_l)E^k_l.
\ee

The (complexified) tangent space $\text{T}_\zx\cO\ot\C$ is spanned
by those $E^k_l$ for which $\zl_k-\zl_l\ne 0$, the space
$D_R(\zx)\ot\C$ is spanned by those $E^k_l$ for which
$\zl_k+\zl_l\ne 0$, the space $D_0(\zx)\ot\C$ is spanned by those
$E^k_l$ for which $\zl_k^2-\zl_l^2\ne 0$, and the distribution
$\cN$ mentioned in the proof of the theorem is spanned by $E^k_l$
for which $\zl_k-\zl_l>0$. The complexified symplectic form reads
$$
\zh^\cO_\zx(i(\zl_k-\zl_l)E^k_l,i(\zl_r-\zl_s)E^r_s)= \la
i(\zl_k-\zl_l)E^k_l,E^r_s\ra_{gl}=-i(\zl_k-\zl_l)\frac{1}{2}(\zd_l^r\zd_s^k),
$$
i.e.
\be
\zh^\cO_\zx(E^k_l,E^r_s)=\frac{1}{2i(\zl_r-\zl_s)}(\zd_l^r\zd_s^k),
\ee
and the complexified Riemannian form
\be
\zg^\cO_\zx(E^k_l,E^r_s)= \zh^\cO_\zx(E^k_l,\cJ_\zx(E^r_s))=
\frac{1}{2\vert\zl_r-\zl_s\vert}(\zd_k^r\zd_s^l).
\ee
As a basis in $u^*(n)$ let us take
\be A^k_l=E^k_l+E^l_k,\quad k\le l,\quad B^k_l=iE^k_l-iE^l_k,\quad k<l.
\ee
It is easy to see that this is an orthonormal basis and that
\be\cJ_\zx(A^k_l)=sgn(\zl_k-\zl_l)B^k_l,\quad\cJ_\zx(B^k_l)=
sgn(\zl_l-\zl_k)A^k_l.
\ee
and
\be\cR_\zx(A^k_l)=sgn(\zl_k+\zl_l)A^k_l,\quad\cJ_\zx(B^k_l)=
sgn(\zl_l+\zl_k)B^k_l.
\ee

Moreover
\be\zh_\zx^\cO(B^k_l,A^r_s)=\frac{\zd^k_r\zd^l_s}{(\zl_k-\zl_l)},\quad
\zh_\zx^\cO(B^k_l,B^r_s)=\zh_\zx^\cO(A^k_l,A^r_s)=0,\quad
\zl_k-\zl_l,\zl_r-\zl_s\ne 0
\ee
and
\be\zg_\zx^\cO(B^k_l,A^r_s)=0,\quad
\zg_\zx^\cO(B^k_l,B^r_s)=\zg_\zx^\cO(A^k_l,A^r_s)=
\frac{\zd^k_r\zd^l_s}{\vert\zl_k-\zl_l\vert},\quad
\zl_k-\zl_l,\zl_r-\zl_s\ne 0.
\ee
In other words \be\zh_\zx^\cO=\sum_{\zl_k-\zl_l\ne
0}\frac{1}{(\zl_k-\zl_l)}\cdot \D b^k_l\we\D a^k_l,\quad
\ee
and \be\zg_\zx^\cO=\sum_{\zl_k-\zl_l\ne
0}\frac{1}{\vert\zl_k-\zl_l\vert} (\D b^k_l\ot\D b^k_l+\D
a^k_l\ot\D a^k_l),
\ee
where
$$b^k_l=\la B^k_l,\cdot\ra_{u^*},\quad a^k_l=\la A^k_l,\cdot\ra_{u^*}$$
are coordinates on $u^*(n)$ such that $B^k_l=\pa_{b^k_l},
A^k_l=\pa_{a^k_l}$. The reduction of the symplectic form $\zh^\cO$
to the leaves of the foliation $\cF_0$
\be(\zh_\zx^\cO)_{\mid\cF_0}=\sum_{\zl_k^2-\zl_l^2\ne 0}\frac{1}{(\zl_k-\zl_l)}\cdot
\D b^k_l\we\D a^k_l,\quad
\ee
is clearly non-degenerate and constitutes, together with the
reduced Riemannian structure
\be(\zg_\zx^\cO)_{\mid\cF_0}=\sum_{\zl_k^2-\zl_l^2\ne 0}\frac{1}{\vert\zl_k-\zl_l\vert}
(\D b^k_l\ot\D b^k_l+\D a^k_l\ot\D a^k_l),
\ee
a K\"ahler structure.

\bigskip\noindent
{\bf Remark.} Of course, when $\zx$ is a projector, then
$\zl_k=1,0$, so $\zl_k-\zl_l\ne
0\Rightarrow\vert\zl_k-\zl_l\vert=1$ and $\zg_\zx^\cO$ reduces to
the canonical scalar product. Note also that the leaves of
$\cF_\zL$ and $\cF_0$ through $\zx$ coincide, except for the rare
case when there are $\zl,\zl'\ne 0$ in the spectrum of $\zx$ such
that $\zl+\zl'=0$. In particular, the foliations $\cF_\zL$ and
$\cF_0$ coincide when reduced to the subset $\cP(\cH)$ of
non-negative operators or to the set $\cD(\cH)$ of density states.
On such leaves the product structure $\cR$ is trivially the
identity.

\section{Composite systems and separability}
Suppose now that our Hilbert space has a fixed decomposition into
the tensor product of two Hilbert spaces $\cH=\cH^1\ot\cH^2$. This
additional input is crucial in studying composite quantum systems
and it has a great impact on the geometrical structures we have
considered. The rest of this paper will be devoted to related
problems.

Observe first that the tensor product map
\be\bigotimes:\cH^1\ti\cH^2\raa\cH=\cH^1\ot\cH^2
\ee
associates the product of rays with a ray, so it induces a
canonical imbedding on the level of complex projective spaces
\bea \text{Seg}:P\cH^1\ti P\cH^2&\raa& P\cH=P(\cH^1\ot\cH^2),\\
(\mid\!x^1\ra\la x^1\!\!\mid,\mid\!x^2\ra\la x^2\!\!\mid)
&\mapsto& \mid\!x^1\ot x^2\ra\la x^1\ot x^2\!\!\mid.
\eea
This imbedding of product of complex projective spaces into the
projective space of the tensor product is called in the literature
the {\it Segre imbedding} \cite{segre}. The elements of the image
$\text{Seg}(P\cH^1\ti P\cH^2)$ in $P\cH=P(\cH^1\ot\cH^2)$ are
called {\it separable} pure states (with respect to the
decomposition $\cH=\cH^1\ot\cH^2$).

The Segre imbedding is related to the (external) tensor product of
the basic representations of the unitary groups $U(\cH^1)$ and
$U(\cH^2)$, i.e. with the representation of the direct product
group in $\cH=\cH^1\ot\cH^2$,
\beas U(\cH^1)\ti U(\cH^2)\ni(\zr^1,\zr^2)&\mapsto& \zr^1\ot\zr^2\in
U(\cH)=U(\cH^1\ot\cH^2),\\
(\zr^1\ot\zr^2)(x^1\ot x^2)&=&\zr^1(x^1)\ot\zr^2(x^2).
\eeas
Note that $\zr^1\ot\zr^2$ is unitary, since the Hermitian product
in $\cH$ is related to the Hermitian products in $\cH^1$ and
$\cH^2$ by \be\label{sp1} \la x^1\ot x^2,y^1\ot y^2\ra_{\cH}=\la
x^1,y^1\ra_{\cH^1}\cdot \la x^2,y^2\ra_{\cH^2}.
\ee
The above group imbedding which, with some abuse of notation, we
will denote by
$$\text{Seg}:U(\cH^1)\ti U(\cH^2)\raa U(\cH),$$
gives rise to the corresponding imbedding of Lie algebras
$$\text{Seg}:u(\cH^1)\ti u(\cH^2)\raa u(\cH),$$
or, by our identification, of their duals
\be\label{imb}\text{Seg}:u^*(\cH^1)\ti u^*(\cH^2)\raa u^*(\cH).
\ee
The original Segre imbedding is just the latter map reduced to
pure states. In fact, a more general result holds true.
\begin{prop} The imbedding (\ref{imb}) maps
$\cP^k(\cH^1)\ti\cP^l(\cH^2)$ into $\cP^{kl}(\cH^1\ot\cH^2)$ and
$\cD^k(\cH^1)\ti\cD^l(\cH^2)$ into $\cD^{kl}(\cH^1\ot\cH^2)$.
\end{prop}
{\it Proof.} Let us take $A^1\in\cP^k(\cH^1)$ and
$A^2\in\cP^l(\cH^2)$. Using bases of eigenvectors $(e^1_i)$ of
$A^1$ and $(e^2_j)$ of $A^2$ to construct a basis $(e^1_i\ot
e^2_j)$ of eigenvectors of $A^1\ot A^2$, one easily sees that the
elements of the spectrum of $A^1\ot A^2$ (counted with
multiplicities) are $\zl_i\zl'_j$, where $A^1(e^1_i)=\zl_i e^1_i$
and $A^2(e^2_j)=\zl'_je^2_j$, so that $A^1\ot
A^2=\text{Seg}(A^1,A^2)$ is non-negatively defined and has rank
$kl$. If $A^1,A^2$ have trace 1, i.e. $\sum_i\zl_i=1$ and
$\sum_j\zl'_j=1$, then $\sum_{i,j}\zl_i\zl'_j=\sum_i\zl_i \cdot
\sum_j\zl'_j=1$. \ep

\medskip\noindent
Let us denote the image $\text{Seg}(\cD^k(\cH^1)\ti\cD^l(\cH^2))$
by $\cS^{k,l}(\cH^1\ot\cH^2)$, the set $\cS^{1,1}(\cH^1\ot\cH^2)$
of {\it separable pure states} simply by $\cS^1(\cH^1\ot\cH^2)$,
and the convex hull
$$\text{conv}\left(\text{Seg}\left(\cD(\cH^1)\ti\cD(\cH^2)\right)\right)$$
of the subset $\text{Seg}\left(\cD(\cH^1)\ti\cD(\cH^2)\right)$ of
all {\it separable states} in $u^*(\cH)$  by $\cS(\cH^1\ot\cH^2)$.
The states from
$$\cE(\cH^1\ot\cH^2)=\cD(\cH^1\ot\cH^2)\setminus\cS(\cH^1\ot\cH^2),$$
i.e. those which are not separable, are called {\it entangled
states}.
\begin{prop} The convex set $\cS(\cH^1\ot\cH^2)$ of separable states is
the convex hull of the set $\cS^1(\cH^1\ot\cH^2)$ of separable
pure states and $\cS^1(\cH^1\ot\cH^2)$ is exactly the set of
extremal points of $\cS(\cH^1\ot\cH^2)$. Moreover,
$\cS^1(\cH^1\ot\cH^2)$, thus $\cS(\cH^1\ot\cH^2)$, is invariant
with respect to the canonical $U({\cal H}^1)\times U({\cal H}^2)$-action on
$u^*(\cH^1\ot\cH^2)$,
$$(T_1,T_2)A=(T_1\ot T_2)\circ A\circ(T_1\ot T_2)^\dag.$$
\end{prop}
{\it Proof.} Let us start with showing that the convex hull of
$\cS^1(\cH^1\ot\cH^2)$ contains
$\text{Seg}\left(\cD(\cH^1)\ti\cD(\cH^2)\right)$ thus equals
$\cS(\cH^1\ot\cH^2)$. Indeed $\cD^1(\cH^i)$ is the set of extreme
points of $\cD(\cH^i)$, $i=1,2$, so that any $A^i\in\cD(\cH^i)$ is
a convex combination $A^i=t_s^i\zr_s^i$ of elements
$\zr^i_s\in\cD^1(\cH^i)$, $i=1,2$. Hence, $A^1\ot A^2$ is the
convex combination
$$A^1\ot A^2=\sum_{s,s'}t^1_st^2_{s'}\cdot\zr^1_s\ot\zr^2_{s'}.$$
On the other hand, every state $\zr^1\ot\zr^2$,
$\zr^i\in\cD(\cH^i)$, $i=1,2$, is in $\cD^1(\cH^1\ot\cH^2)$, i.e.
it is an extremal point of $\cD(\cH^1\ot\cH^2)$. Therefore it
cannot be written as a non-trivial convex combination of elements
from $\cD(\cH^1\ot\cH^2)$, thus from a smaller set
$\cS(\cH^1\ot\cH^2)$. The invariance is obvious, since
$$(T_1\ot T_2)\circ(\zr_1\ot\zr_2)\circ(T_1^\dag\ot T_2)^\dag=
(T_1\zr_1T_2^\dag)\ot(T_1\zr_2T_2^\dag)$$ and
$(T_1\zr_iT_2^\dag)\in\cD^1(\cH^i)$ for $\zr_i\in\cD^1(\cH^i)$.
\ep

\medskip\noindent
Since we are working in a finite-dimensional space, the closeness
of the corresponding hulls is automatic that can be derived from
the following lemma.
\begin{lem} If $V$ is an $n$-dimensional real vector space and $x$
is a convex combination $x=\sum_{i=1}^mt_ix_i$ of certain points
of $V$, then $x$ is a convex combination of at most $(n+1)$ points
among $x_i$'s.
\end{lem}
{\it Proof.} It suffices to prove that $x$ is a convex combination
of $(m-1)$ of $x_i$'s, provided $m>n+1$. Of course, we can assume
that all $t_i>0$. If $m>n+1$, then there are $a_i\in\R$, not all
equal 0, such that $\sum_1^m a_i=0$ and $\sum_1^ma_ix_i=0$. There
is $i_0$ such that $\vert a_{i_0}/t_{i_0}\vert$ is maximal among
$\vert a_{i_0}/t_{i_0}\vert$, $i=1,\dots,m$. We can assume without
loss of generality that $i_0=m$. Hence
$$x=\sum_{i=1}^{m-1}\left(t_i-\frac{a_it_m}{a_m}\right)x_i
$$
and the above combination is convex, since
$(t_i-\frac{a_it_m}{a_m})\ge 0$ and
$$\sum_1^{m-1}(t_i-\frac{a_it_m}{a_m})=\sum_1^m(t_i-\frac{a_it_m}{a_m})
=\sum_1^mt_i=1.$$ \ep
\begin{prop}\label{compact} The convex hull $\text{conv}(E)$ of a compact subset
$E$ in a finite dimensional real vector space $V$ is compact.
\end{prop}
{\it Proof.} Suppose that the dimension of the space is $n$ and
denote by $\zD_{n+1}$ the compact $(n+1)$-dimensional simplex
$$\zD_{n+1}=\{ t=(t_1,\dots,t_{n+1}):t_i\ge 0,\sum_1^{n+1}t_i=1\}.
$$
According to the above lemma, $\text{conv}(E)$ is the image of the
compact set $\zD\ti E\ti\dots\ti E$ ($E$ appears in the product
$(n+1)$-times) under the continuous map
$$\zD\ti E\ti\dots\ti E\ni(t,x_1,\dots,x_{n+1})\mapsto\sum_1^{n+1}
t_ix_i\in V.
$$
\ep
\begin{cor} The set $\cS(\cH^1\ot\cH^2)$ is a compact subset of
$u^*(\cH^1\ot\cH^2)$.
\end{cor}

\medskip\noindent
The entangled states play an important role in quantum computing
and one of main problems is to decide effectively whether a given
composite state is entangled or not. An abstract measurement of
entanglement can be based on the following observation (see also
Ref. \cite{Vidal})

\smallskip
Let $E$ be the set of all extreme points of a compact convex set
$K$ in a finite-dimensional real vector space $V$ and let $E_0$ be
a compact subset of $E$ with the convex hull
$K_0=\text{conv}(E_0)\subset K$. For every non-negative function
$f:E\raa\R_+$  define its extension $f_K:K\raa\R_+$ by
\be\label{cf}f_K(x)=\inf_{x=\sum t_i\za_i}\sum t_if(\za_i),
\ee
where the {\it infimum} is taken with respect to all expressions
of $x$ in the form of convex combinations of points from $E$.
Recall that that, according to Krein-Milman theorem, $K$ is the
convex hull of its extreme points.
\begin{theo}\label{hull}
For every non-negative continuous function $f:E\raa\R_+$ which
vanishes exactly on $E_0$ the function $f_K$ is convex on $K$ and
vanishes exactly on $K_0$
\end{theo}
{\it Proof.} It is completely obvious that $f_K$ vanishes on the
convex hull of $E_0$. The function $f_K$ is convex, since for
every convex combination $x=t_iy_i$ of points of $K$ and every
$\ze>0$ we can find extreme points $\za_j$ with convex
combinations $y_i=s_i^j\za_j$ and $f_K(y_i)>s_i^jf(\za_j)-\ze$.
Hence
$$f_K(t_iy_i)=f_K(t_is^j_i\za_j)\le t_is^j_if(\za_j)<t_i(f(y_i)+\ze)
=t_if_K(y_i)+\ze.$$ Due to arbitrariness of $\ze>0$ we get
$$f_K(t_iy_i)\le t_if_K(y_i).$$
Note finally that $f_K$ vanishes exactly on $K_0$. Indeed $K_0$ is
compact due to proposition \ref{compact} and if $x\notin K_0$,
then $x$ and $K_0$ can be separated by a hyperplane, i.e. there is
a linear functional $\zf:V\raa\R$ such that $\zf(x)=a>0$ and $\zf$
is negative on $K_0$. Denote by $E_1$ the (compact) set of those
points from $E$ on which $\zf$ takes non-negative values and by
$F$ the minimum of $f$ on $E_1$. Of course, $F>0$, since $E_1\cap
E_0=\emptyset$. Let $M\in\R$ be the maximum of $\zf$ on $E$. Of
course, $M>0$. For any realization $x=t_i\za_i$ of $x$ as a convex
combination of points of $E$ we have
$$a=\zf(x)=\sum_i t_i\zf(\za_i)\le\sum_{\za_i\in
E_1}t_i\zf(\za_i)\le M\sum_{\za_i\in E_1}t_i.$$ On the other hand,
$$\sum_i t_if(\za_i)\ge\sum_{\za_i\in E_1}t_if(\za_i)\ge
F\sum_{\za_i\in E_1}t_i\ge \frac{aF}{M},$$ so
$f_K(x)\ge\frac{aF}{M}>0$. \ep
\begin{cor}
Let $F:\cD^1(\cH^1\ot\cH^2)\raa\R_+$ be a continuous function
which vanishes exactly on $\cS^1(\cH^1\ot\cH^2)$. Then
$$\zm=F_{\cD(\cH^1\ot\cH^2)}:\cD(\cH^1\ot\cH^2)\raa\R_+$$
is a measure of entanglement, i.e. $\zm$ is convex and
$\zm(x)=0\Leftrightarrow x\in\cS(\cH^1\ot\cH^2)$. Moreover, if $f$
is taken $U(\cH^1)\ti U(\cH^2)$-invariant, then $\zm$ is $U(\cH^1)\ti
U(\cH^2)$-invariant.
\end{cor}
{\it Proof.} The first part is a direct consequence of Theorem
\ref{hull}. Also the invariance of $\zm$ is clear:
$$\zm(T\zr T^\dag)=\inf_W(t_if(\za_i))=\inf_{W'}(t_if(T\za_iT^\dag))=
\inf_{W'}(t_if(\za_i))=\zm(\zr),$$ where $T$ is in the
corresponding group,
$$W=\{(t_i,\za_i):T\zr T^\dag=\sum t_i\za_i,\
\za_i\in\cS^1(\cH^1\ot\cH^2), t_i\ge 0,\sum t_i=1\},
$$
and
$$W'=\{(t_i,\za_i):\zr=\sum t_i\za_i,\
\za_i\in\cS^1(\cH^1\ot\cH^2), t_i\ge 0,\sum t_i=1\}.
$$
\ep

\medskip\noindent
A careful study of the geometry of $u^*((\cH^1\ot\cH^2)$ and
criteria of entanglement we postpone to a separate paper.

\section{Acknowledgements}
We would like to thank V.~S.~Varadarajan for useful discussions on
the contents of this paper. This work was supported by the Polish
Ministry of Scientific Research and Information Technology under
the (solicited) grant No PBZ-Min-008/P03/03 and partially
supported by PRIN SINTESI.

%\noindent Janusz GRABOWSKI\\Polish Academy of Sciences\\Institute
%of Mathematics\\\'Sniadeckich 8\\P.O. Box 21\\00-956 Warsaw,
%Poland\\Email: jagrab@impan.gov.pl\\\\

\begin{thebibliography}{99}

\bibitem{Dirac}
P.~A.~M.~Dirac, The Principles of Quantum Mechanics, 4th edition
(Pergamon, Oxford, 1958)

\bibitem {VonNeumann} J.~von~Neumann, Foundations of Quantum
Mechanics, Princeton Univ.\ Press, Princeton 1971

\bibitem{Pauli} W.~Pauli, General principles of quantum mechanics.
Translated from the German by P.~Achuthan and K.~Venkatesan,
Springer-Verlag, Berlin-New York, 1980

\bibitem{Weyl} H.~Weyl, The theory of groups and quantum
Mechanics, Dover Publ., New York 1931

\bibitem{Wigner} E.~Wigner, On unitary representations of the inhomogeneous
Lorentz group, Ann.\ of Math.\  \textbf{40} (1939), no.\ 1, 149--204

\bibitem{Wigner1} E.~P.~Wigner, Group theory and its application to the
quantum mechanics of atomic spectra, Expanded and improved ed.
Translated from the German by J.~J.~Griffin. Pure and Applied
Physics. Vol.\ 5 Academic Press, New York-London 1959

\bibitem{Carinena}  J.~F.~Cari\~nena, J.~Grabowski, G.~Marmo, Lie--Scheffers
Systems: A Geometrical Approach (Bibliopolis, Napoli,
2000)\newline J.~F.~Cari\~nena, G.~Marmo, J.~Nasarre, The nonlinear
superposition principle and the Weyl--Norman method, Int.\ J. Mod.\
Phys.\ \textbf{A 13} (1998) 3601--362

\bibitem{MMSZ}  V.~I.~Man'ko, G.~Marmo, E.~C.~G.~Sudarshan and F.~Zaccaria,
Inner composition law of pure states, Phys.\ Lett.\ \textbf{A 273}
(2000) 31--36

\bibitem{Cirelli}  R.~Cirelli, M.~Gatti, A.~Mani\'{a}, On the
nonlinear extension of quantum superposition and uncertainty
principles, J. Geom.\ and Phys.\ \textbf{29} (1999) 54--86

\bibitem{Jordan}  P.~Jordan, J.~von~Neumann, E.~Wigner, On an algebraic
generalization of the quantum mechanical formalism, Ann.\ Math.\
\textbf{35} (1934) 29--54

\bibitem{Cantoni} V.~Cantoni, The Riemannian structure on the space of quantum-like
systems, Comm.\ Math.\ Phys.\ \textbf{55} (1977) 189--193; Intrinsic
geometry of the quantum-mechanical phase space, Hamiltonian
systems and correspondence
principle, Licei.\ Rend.\ Sc.\ Fis.\ Mat.\ e Nat.\ \textbf{LXII} (1977) 628--636%

\bibitem{all}  R.~Cirelli, P.~Lanzavecchia, A.~Mani\'{a}, Normal pure states
of the von Neumann algebra of bounded operators as K\"{a}hler manifold, J.
Phys.\ A: Math.\ Gen.\ \textbf{15} (1983) 3829--3835
\newline R.~Cirelli, P.~Lanzavecchia, Hamiltonian vector fields in quantum
mechanics, Il.\ Nuovo Cimento \textbf{B 79} (1984) 271--283
\newline M.~C.~Abbati, R.~Cirelli, P.~Lanzavecchia, A.~Mani\'{a},
Pure states of general quantum-mechanical systems as K\"{a}hler Bundle, Il.\ Nuovo Cimento \textbf{B 83} (1984) 43--60
\newline A.~Bloch, An infinite-dimensional Hamiltonian system on a projective
Hilbert space, Transactions of the Am.\ Math.\ Soc.\ \textbf{302} (1987) 787--796
\newline A.~Heslot, Quantum mechanics as a classical theory, Phys.\ Rev.\ \textbf{D 31} (1985) 1341--1348
\newline D.~J.~Rowe, A.~Ryman, G.~Rosensteel, Many-body quantum mechanics as a
symplectic dynamical system, Phys.\ Rev.\ \textbf{A 22} (1980) 2362--2373
\newline T.~R.~Field, J.~S.~Anandan, Geometric phases and coherent states,
J. Geom.\ Phys.\ \textbf{50} (2004) 56--78
\newline D.~C.~Brody, L.~P.~Hughston, Geometric quantum mechanics,
J. Geom.\ Phys.\ \textbf{38} (2001) 19--53
\newline F.~Strocchi, Complex coordinates and quantum mechanics,
Rev.\ Mod.\ Phys.\ \textbf{38} (1956) 36--40
\newline A.~Ashtekar, T.~A.~Shilling, Geometrical
formulation of quantum mechanics, gr-qc/9706069, in On Einstein's
Path, A.~Harvey (Ed.) Springer-Verlag, Berlin (1998)
\newline V.~I.~Ma\'nko, G.~Marmo, E.~C.~G.~Sudarshan, and F.~Zaccaria
The Geometry of Density States, Rep.\ Math.\ Phys.\ \textbf{55} (2005) 405--422

\bibitem{Fano} U.~Fano, Description of states in quantum mechanics
by density matrix and operator techniques, Rev.\ Mod.\ Phys.\ \textbf{29} (1957) 74--93

\bibitem{KKS} B.~Kostant, Quantization and Unitary Representations. Part I.
Prequantization, in: Lecture Notes in Mathematics 170. Springer-Verlag,
Berlin 1970
\newline A.~A.~Kirillov, Elements of the Theory of Representations,
Springer-Verlag, Berlin 1976
\newline J-M.Souriau, Structure des Systemes Dynamiques, Dunod, Paris 1970

\bibitem{Sommers} H.--J.~Sommers, K.~\.Zyczkowski, Bures volume of
the set of mixed quantum states, J. Phys.\ A \textbf{36} (2003) 10083--10100
\newline K.~\.Zyczkowski, H.--J.~Sommers, Hilbert--Schmidt volume of the set
of mixed quantum states, J.~Phys.\ A \textbf{36} (2003) 10115-10130

\bibitem{Ch-Jam}  D.~Chru\'sci\'nski, A.~Jamio\l kowski, Geometric Phases in
Classical and Quantum Mechanics (2004) Birkh\"{a}user, Boston

\bibitem{phase} M.~V.~Berry, Quantal phase factors accompanying adiabatic
changes, Proc.\ Roy.\ Soc.\ London A {\bf 392} (1984) 45--57
\newline N.~Mukunda and R.~Simon, Quantum Kinematic approach to the geometric
phase. I.General formalism, Ann.\ Phys.\ {\bf 228} (1993) 205--268
\newline G.~Esposito, G.~Marmo, E.~C.~G.~Sudarshan, From Classical to Quantum
Mechanics, Cambridge University Press, Cambridge 2004

\bibitem{dittmann:92} J.~Dittmann and G.~Rudolph, On a connection governing
parallel transport along $2\times 2$ density matrices, J.\ Geom.\
Phys.\ {\bf 10} (1992) 93-106

\bibitem{schlienz:95} J.~Schlienz and G.~Mahler, Description of entanglement,
Phys.\ Rev.\ A {\bf 52} (1995) 4396-4404

\bibitem{grassl:98} M.~Grassl, M.~R\"otteler, and T.~Beth, Computing local
invariants of quantum-bit systems, Phys.\ Rev.\ A {\bf 58} (1998) 1833--1839

\bibitem{makhlin:02} Y.~Makhlin, Locally invariant properties of two-qubit
gates and mixed states and optimization of quantum computations, Quant.\ Info.\ Proc.\ {\bf 1} (2002) 243-252

\bibitem{briand:03} E.~Briand, J.--G.~Luque and J.--Y.~Thibon, A complete set of
covariants of the four qubit system, J. Phys.\ A: Math.\ Gen.\ {\bf 36} (2003)
9915-9927

\bibitem{luque:03} J.--G.~Luque and J.--Y.~Thibon, Polynomial invariants of four
qubits, Phys.\ Rev.\ A {\bf 67} (2003) 042303

\bibitem{linden:98} N.~Linden and S.~Popescu, On multi-particle entanglement,
Fortschr.\ Phys.\ {\bf 46} (1998) 567--578

\bibitem{linden:99} N.~Linden, S.~Popescu, and A.~Sudbery, Non-local
parameters of multi-particle density matrices, Phys.\ Rev.\ Lett.\ {\bf 83}
(1999), 243--247

\bibitem{kus:01} M.~Ku\'s and K.~\.Zyczkowski, Geometry of entangled states,
Phys.\ Rev.\ A {\bf 63} (2001), 032307

\bibitem{sinolecka:02} M.~M.~Sino{\l}\c{e}cka, K.~\.Zyczkowski, and
M.~Ku\'s, Manifolds of equal entanglement for composite quantum system, Acta
Phys.\ Pol.\ B {\bf 33} (2002), 2081--2095

\bibitem{moseri:01} R.~Moseri and R.~Dandoloff, Geometry of entangled states,
Bloch spheres and Hopf fibrations, J. Phys.\ A {\bf 34} (2001), 10243--10252

\bibitem{moseri:03} R.~Moseri, Two and three qubits geometry and Hopf
fibrations, arXiv:quant-ph/0310053, in Topology in Condensed Matter
Series: Springer Series in Solid-State Sciences, Vol.\ 150
Monastyrsky, Michael I. (Ed.), in print

\bibitem{levay:04} P.~L\'evay, The geometry of entanglement: metrics, connections
and the geometric phase, J. Phys.\ A: Math.\ Gen.\ {\bf 37} (2004)
1821--1841

\bibitem{heydari:05} H.~Heydari, Segre variety, conifold, Hopf fibration,
and separable multi-qubit states, arXiv:quant-ph/0506043

\bibitem{heydari:05a} H.~Heydari and G.~Bj\"ork, Complex multi-projective
variety and entanglement, J. Phys.\ A: Math.\ Gen.\ {\bf 38} (2005)
3203--3211

\bibitem{levay:05} P.~L\'evay, On the geometry of a class of $N$-qubit entanglement
monotones, arXiv:quant-ph/0507070

\bibitem{heydari:05b} H.~Heydari, Entanglement monotone for multi-qubit states
based on geometric invariant theory, arXiv:quant-ph/0507077

\bibitem{levay:05a} P.~L\'evay, Geometry of three-qubit
entanglement, Phys.\ Rev.\ A {\bf 71} (2005) 012334

\bibitem{chen:01} H.~Chen, Quantum Entanglement and Geometry of
Determinantal Varieties, arXiv:quant-ph/0110103

\bibitem{shi:01} M.~Shi and J.~Du, Boundary of the Set of Separable States,
arXiv:quant-ph/0103016

\bibitem{Pizzocdeura}  R.~Cirelli, A.~Mani\'{a}, L.~Pizzocchero, Quantum
mechanics as an infinite-dimensional Hamiltonian system with uncertainty
structure. Int.\ J. Math.\ Phys.\ \textbf{31} (1990) 2891--2897;
II. J. Math.\ Phys.\ \textbf{31} (1990) 2898--2903

\bibitem{Benvegnu}  A.~Benvegn\'{u}, N.~Sansonetto, M.~Spera, Remarks on
Geometrical Quantum Mechanics, J. Geom.\ Phys.\ \textbf{51} (2004)
229--243

\bibitem{Kibble} T.~W.~Kibble, Geometrization of Quantum
Mechanics, Comm.\ Math.\ Phys.\ \textbf{65} (1979) 189--201

\bibitem{segre} R.~Hartshorne, Algebraic geometry, Springer, 1977, Sect. IV.2.

\bibitem{Vidal} G.~Vidal, Entanglement monotones, J. Mod.\ Opt.\ \textbf{47}
(2000) 355--376

\end{thebibliography}
\end{document}